\documentclass[twocolumn]{aastex63}

\accepted{September 15, 2020}
\submitjournal{ApJ}

\shorttitle{Spatial variations in the solar emission}
\shortauthors{Sharma et al.}
\graphicspath{{./}{figures/}}

\begin{document}

\title{Propagation Effects in Quiet Sun Observations at Meter Wavelengths}

\correspondingauthor{Rohit Sharma}
\email{rohit.sharma@fhnw.ch, rohitcbscient@gmail.com}

\author[0000-0003-0485-7098]{Rohit Sharma}
\affil{Fachhochschule Nordwestschweiz \\
Bahnhofstrasse 6, \\
5210 Windisch, Switzerland}

\author[0000-0002-4768-9058]{Divya Oberoi}
\affil{National Centre for Radio Astrophysics \\ 
Tata Institute of Fundamental Physics, S. P. Pune University, \\ Ganeshkhind, Pune 411007, India}



\begin{abstract}

Quiet sun meterwave emission arises from thermal bremsstrahlung in the MK corona, and can potentially be a rich source of coronal diagnostics. 
On its way to the observer, it gets modified substantially due to the propagation effects - primarily refraction and scattering - through the magnetized and turbulent coronal medium, leading to the redistribution of the intensity in the image plane.
By comparing the full-disk meterwave solar maps during a quiet solar period and the modelled thermal bremsstrahlung emission, we characterise these propagation effects. 
The solar radio maps between 100 and 240 MHz come from the Murchison Widefield Array. FORWARD package is used to simulate thermal bremsstrahlung images using the self-consistent Magnetohydrodynamic Algorithm outside a Sphere coronal model. 
The FORWARD model does not include propagation effects.
The differences between the observed and modelled maps are interpreted to arise due to scattering and refraction.
There is a good general correspondence between the predicted and observed brightness distributions, though significant differences are also observed.
We find clear evidence for the presence of significant propagation effects, including anisotropic scattering.
The observed radio size of the Sun is 25--30\% larger in area.
The emission peak corresponding to the only visible active region shifts by 8'--11' and its size increases by 35--40\%.
Our simple models suggest that the fraction of scattered flux density is always larger than a few tens of percent, and varies significantly between different regions.
We estimate density inhomogeneities to be in the range 1--10\%.

\end{abstract}


\section{Introduction} \label{sec:intro}

The solar emission at low radio frequencies originates in the million K corona.
The thermal bremsstrahlung from the coronal plasma gives rise to the relatively steady and spectrally smooth, quiet Sun emission.
At meterwaves (100--300 MHz), this emission arises from coronal heights spanning $\sim$1.1-1.5 $R_{\odot}$, where $R_{\odot}$ is the solar radius \citep[e.g][]{Newkirk1961}, a region of rapid drop in coronal density.
As this radiation traverses the coronal medium on its way to the Earthbound observer,  
the coronal optical depth, $\tau$, along typical ray paths is neither low enough to be approximated as optically thin, nor high enough to be treated as optically thick \citep[e.g.][]{Kundu1965,McLean1985,Dulk1974,Alissandrakis1985,Alissandrakis1994,Mercier2015}.
It is also well known that through much of this region $\beta<1$, where $\beta$ is the ratio of the thermal to magnetic pressure in the plasma \citep{Gary2001}. 
Since the magnetic fields energetically dominate corona, the coronal magnetic structures give rise to corresponding structures in electron density.
Thus superposed on the very large scale purely radial coronal density gradients \citep{Newkirk1961,Saito1977,Zucaa2014}, are smaller length scale density structures arising from diversity of coronal magnetic structures.
The medium is anisotropic but bears a strong radial symmetry due to the presence of large scale radial density gradient, and the mostly radial structure of open magnetic fields. 
The presence of closed magnetic field structures introduces significant departures from this radial symmetry.
The ubiquitous turbulence gives rise to small scale density fluctuations, which give rise to scattering.
 As it is easier for plasma to flow along the magnetic fields than perpendicular to it, it is reasonable to expect the small scale density inhomogenties to be elongated along the local magnetic fields.
This is supported by observational evidence \citep{Anantha1996, Bastian1999, Gotoskar2001,  Kontar2017}.

The propagation through the coronal medium leads to an apparent redistribution of the emission in the image plane.
In addition, for impulsive emissions, it also gives rise to a smearing in time.
The latter is observed in studies of type III solar bursts at very low frequencies \citep[e.g.]{Krupar2018}.
It is usually much smaller at meterwaves and is not discussed here any further.
Accounting for propagation effects through the coronal medium, requires dealing with ray propagation through an anisotropic magnetised plasma with structures spanning a very large range in sizes, and in the strong scattering regime. 
This has remained challenging and only recently have the first simulations which build on prior work \citep[e.g.][]{Steinberg1971, Arzner1999, Thejappa2008} to account for scattering from anisotropic small scale turbulence, and refraction from large scale coronal turbulence begun to appear \citep[e.g.][]{Kontar2019}.

Numerous prior studies have focused on active solar radio emissions, largely from the well known solar bursts from types I through V \citep[e.g.][and many references therein]{Pick2008}.
A large fraction of these studies have relied on data from radio spectrographs which provide no imaging information, and hence could not be used for studying propagation effects.
There have been multiple studies using brightness temperatures derived from solar radio images.
Very few of them, however, take propagation effects into account \citep{Steinberg1971,Thejappa1992,Thejappa1994,Thejappa2008}.

Studies of the quiet solar corona have, however, been far fewer as they are significantly more challenging.
These challenges come from issues like the complex radio morphology of the quiet corona; the comparatively low contrast of quiet time coronal emission features; the large angular scale of the emission which is intrinsically harder to image; along with, till recently, the necessity to make synthesis observations over many hours to arrive at reasonable meterwave coronal images \citep[e.g.][]{Kundu1987, Mercier2009, Mercier2012, Vocks2018}.
These quiet sun studies have already established the large scale correspondences between features observed in optical, UV and X-ray bands with those observed at meterwave radio band.
The solar features observed at optical wavelengths show counterparts in meterwaves at large spatial-temporal scales in synoptic maps \citep[e.g][]{Kundu1987}.
Coronal helmets and active streamers give rise to enhanced thermal emission at low radio frequencies \citep[e.g.][]{Axisa1971}. 
The EUV brighter regions from hot coronal loops and fainter coronal holes both have a meterwave counterparts \citep[e.g][]{Dulk1974,  Lantos1980}.  
Coronal holes, especially many of the ones close to the disk centre have been found to transition from being radio dark at lower coronal heights (higher observing frequencies) to being radio bright at higher coronal heights (lower observing frequencies) \citep{Dulk1977,Lantos1999,Mercier2012,Mercier2009}. 
There have been a few attempts to explain this invoking aspects ranging from orientation of magnetic fields to refractive effects \citep{Alissandrakis1994,Golap1994,Rahman2019}.

Propagation effects are believed to be responsible for giving rise to apparent shifts in the observed locations of compact radio sources \citep{Steinberg1971, Gordovskyy2019, Kontar2019}.  
Scattering is regarded to be responsible for the lower than expected coronal brightness temperature, $T_B$, especially towards the decameter wave band \citep{Thejappa1992, Thejappa1994, Thejappa2008}. 
Multiple studies have used radio observations to estimate the level and nature of coronal density inhomogenities \citep[e.g.][]{Bastian1994, Thejappa2008, Subramanian2011, Ingale2015, Mohan2019}. 
\citet{Mercier2015} have attempted to estimate the electron density and temperature distributions in the solar corona from multi-frequency radio observations.

We aim to compare high quality solar radio images from an exceptionally quiet period with the corresponding thermal bremsstrahlung maps obtained using data driven MHD simulations.
The radio images come from the Murchison Widefield Array (MWA), which represents the state-of-the-art in solar imaging in this band \citep{Mondal2019,Mondal2020b,Mondal2020a,Mohan2019b,Mohan2019}.
The corresponding simulated thermal bremsstrahlung maps are obtained using the FORWARD software \citep{Gibson2016}, which models the radio brightness distribution. 
It uses the state-of-the-art description of the physical state of the corona computed using the Magnetohydrodynamic Algorithm outside a Sphere (MAS) MHD simulation data cubes.
MAS models global 3D coronal magnetic fields using sophisticated force-balance and attempts to provide an accurate representation of the solar corona. 
Among other applications, the FORWARD synthesised radio brightness distributions, based on physical state computed using MAS models, have been used for investigations of the solar radio emissions \citep[e.g.][]{McCauley2017, McCauley2018, McCauley2019, Rahman2019}.
While the observed radio maps, naturally, experience all of the propagation effects in their full glory, the present simulated maps do not include these effects.
The comparison of these maps, thus provides an opportunity for a detailed study of the impact of the propagation effects, and forms the primary objective of this work.
The choice of a quiet period is motivated by the desire to limit the complexity of the problem to a situation of low temporal variability.

The data used is described in Sec. \ref{Sec:data}, and the thermal solar radio emission and the FORWARD maps in Sec. \ref{Sec:thermal-radio}.
Section \ref{Sec:radio-analysis} describes the analysis of the MWA radio data. We also compare the observed and simulated radio maps, along with an exploration of the  MAS model data cube in the same section.
Attempts to quantify the propagation effects are presented in Sec. \ref{Sec:quantifying-scattering}, followed by a discussion and summary in Sec. \ref{Sec:discssion}.

\section{Data}
\label{Sec:data}

Operating in the 80 -- 300 MHz band and located in the radio quiet Western Australia, the MWA is a precursor to the Square Kilometer Array.
MWA is one of the first arrays to implement a `large-N' design, where N is the number of elements \citep{Lonsdale2009,Tingay2013}.
This work uses data from the Phase-I of the MWA, which had 128 elements distributed in a centrally condensed configuration over a $\sim$3 km footprint.
The dense uv sampling, resulting from the 8,128 baselines sampling a uv plane extending only to a few $k\lambda$, makes MWA exceptionally well suited for solar imaging at these frequencies \citep{Mondal2019}. 

\subsection{Data Selection}

\begin{figure*}
\centering
\begin{tabular}{cc}
  
\includegraphics[width=0.8\columnwidth]{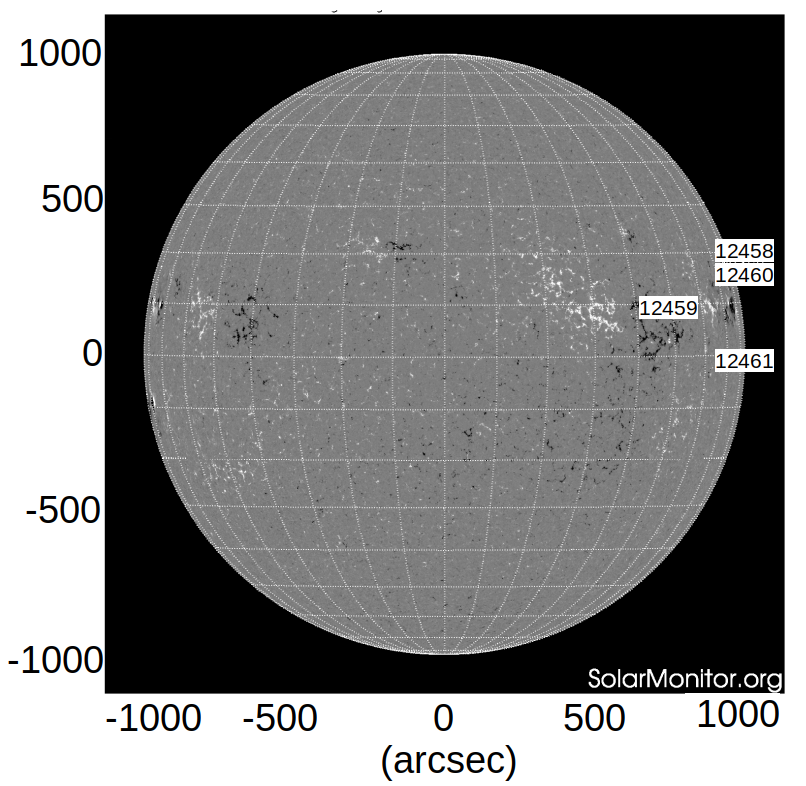}  
& \includegraphics[width=0.8\columnwidth]{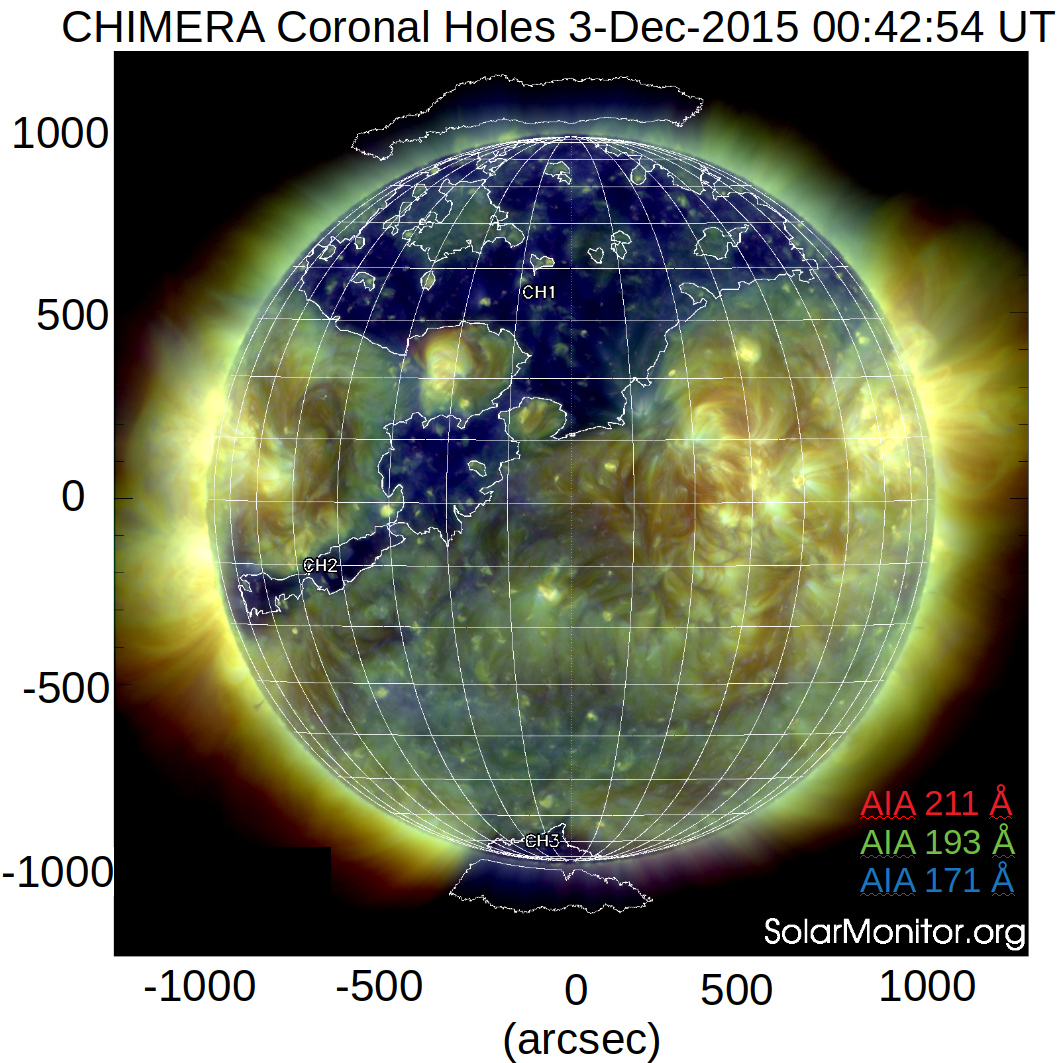} \\
(a) HMI magnetogram  & (b) EUV (AIA)   \\
\end{tabular}
\caption{
Left panel: HMI magnetogram  
Right panel: Composite EUV image for wavelengths 211 (red), 193 (green) and 171 (blue) $\AA$. The white contours show the extent of the coronal hole determined by CHIMERA. 
}  
\label{Fig:aia_hmi}
\end{figure*}

The data used for this study come from a period of very low solar activity.
These data were obtained on 03 December, 2015 as a part of the G0002 solar observing program.
This day is characterized as a period of low solar activity by \textit{solarmoniter.org}.
Extreme UV (EUV) images from the Helioseismic and Magnetic Imager (HMI) instrument, onboard the Solar Dynamic Observatory (SDO) show a smooth featureless disk across much of the photosphere, except for a few flux emerging footprints of active regions near the eastern and western limbs (Fig. \ref{Fig:aia_hmi}).
Four weak bi-polar regions were reported. 
Three of them were located very close to the limb at N11W78, N12W83 and S07W91; while one was located at N05W46.
Only two GOES C-class X-ray flares were reported in the past two days, and on this day two C-class flares were reported from the active region 12458 at 04:25 UT (C2.2) and 06:01 (C1.2) UT respectively.
The composite solar map from Atmospheric Imaging Assembly (AIA) instrument onboard the SDO shows the absence of on-disk active regions and the presence of a large coronal hole (Fig. \ref{Fig:aia_hmi}).
Quantifying the extent of the coronal holes using the  Coronal Hole Identification via Multi-thermal Emission Recognition Algorithm (CHIMERA; \cite{Garton2018}) on AIA multi-thermal images lead to a detection of three coronal holes covering a very substantial $\sim$23.8~\% of total solar disk.

\section{Radio emission from thermal Bremsstrahlung}
\label{Sec:thermal-radio}

The thermal bremsstrahlung emission depends directly on the local plasma parameters like electron density ($n_e$), temperature ($T_e$), and the local magnetic field ($B$). 
The optical depth ($\tau$) of the ray paths through the coronal medium are a significant fraction of unity \citep{McLean1985}.
This, on the one hand, makes radio observations a very interesting probe of the coronal medium as the observed brightness temperature ($T_B$) is an integral over the entire Line-of-Sight (LoS). On the other hand, it significantly increases the complexity of modeling this emission as one is neither firmly in the optically thin or the optically thick regimes and so one cannot make the corresponding simplifying assumptions. 
The corona is not static, its physical state changes continuously with the evolution of the photospheric footpoints of coronal features. 
Realistic modeling of the coronal structures is essential to arrive at correspondingly realistic estimates of the  thermal emission arising from these structures.
These models need to be much more detailed than the smooth large scale coronal $n_e$ models which only accommodate radial or at most also latitudinal $n_e$ variations \citep[e.g.][]{Newkirk1967,Saito1977,Mann1999}. 

\subsection{FORWARD modelled maps}
\label{subsec:fwd_model}

\begin{figure*}
\begin{tabular}{cccc}
\resizebox{42mm}{!}{
\includegraphics[trim={0.0cm 0cm 0.0cm 0.0cm},clip,scale=0.6]{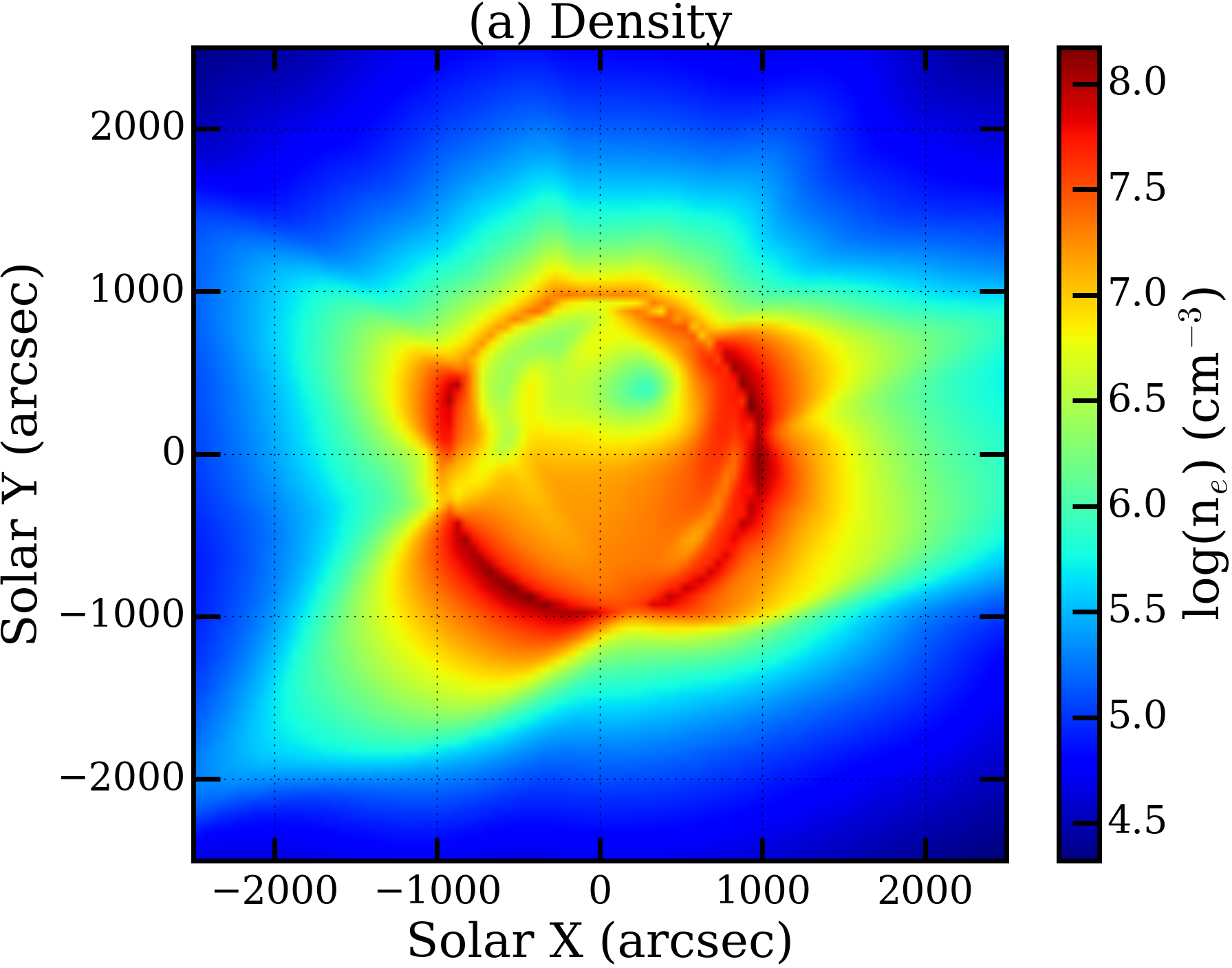}}
     &  
\resizebox{42mm}{!}{
\includegraphics[trim={0.0cm 0cm 0.0cm 0.0cm},clip,scale=0.8]{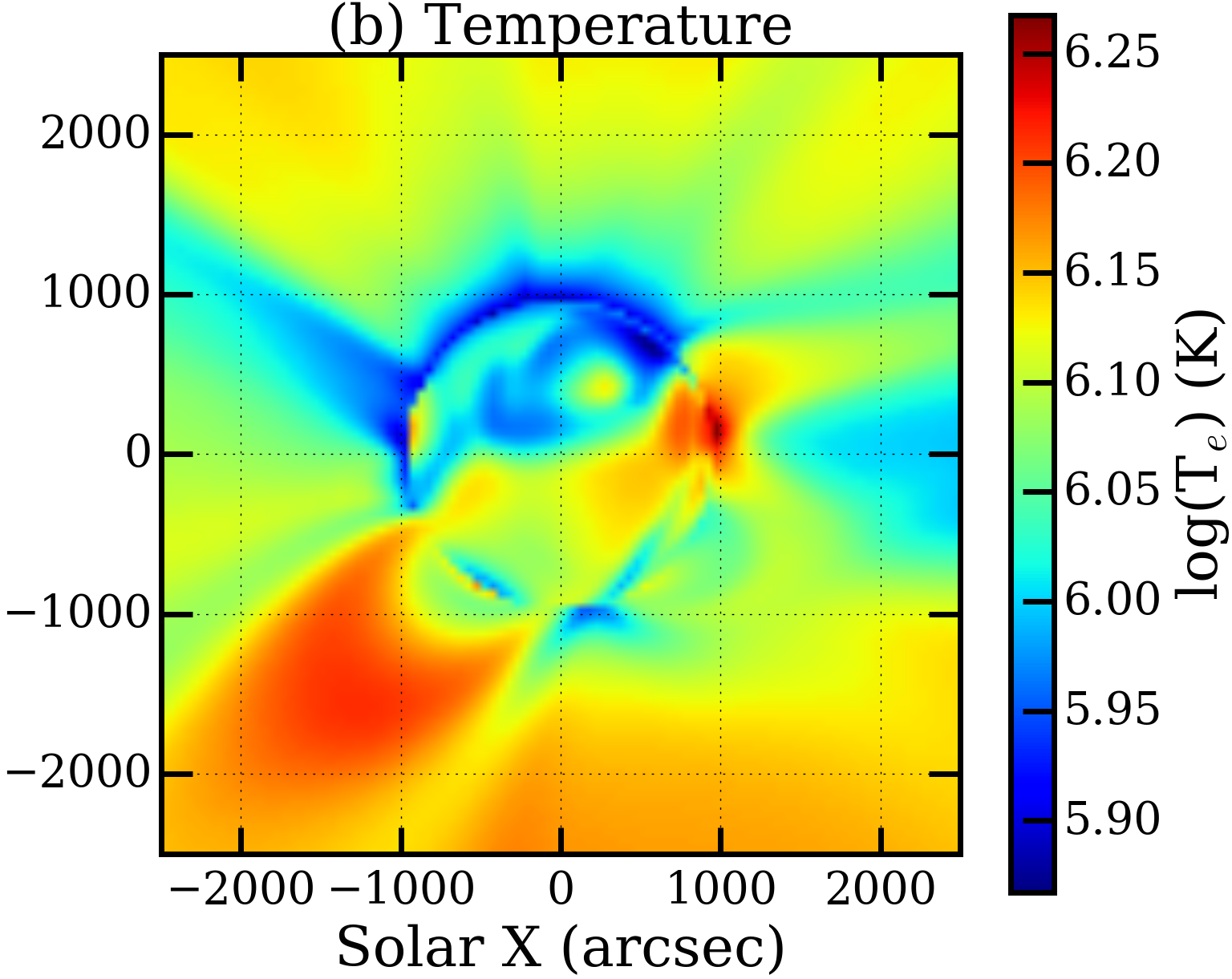}}
    &
\resizebox{42mm}{!}{
\includegraphics[trim={0.0cm 0cm 0.0cm 0.0cm},clip,scale=0.8]{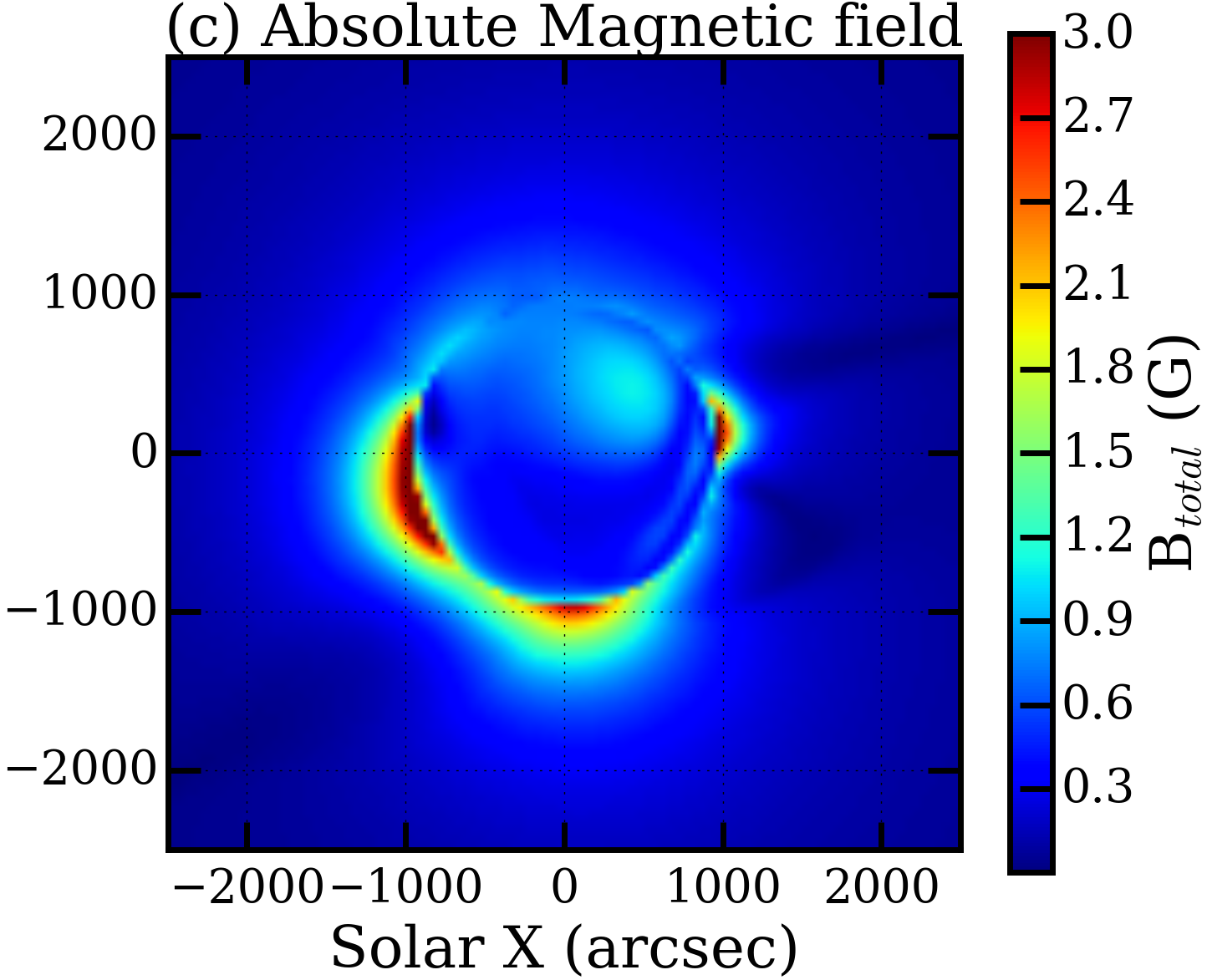}} 
   &
\resizebox{42mm}{!}{
\includegraphics[trim={0.0cm 0cm 0.0cm 0.0cm},clip,scale=0.8]{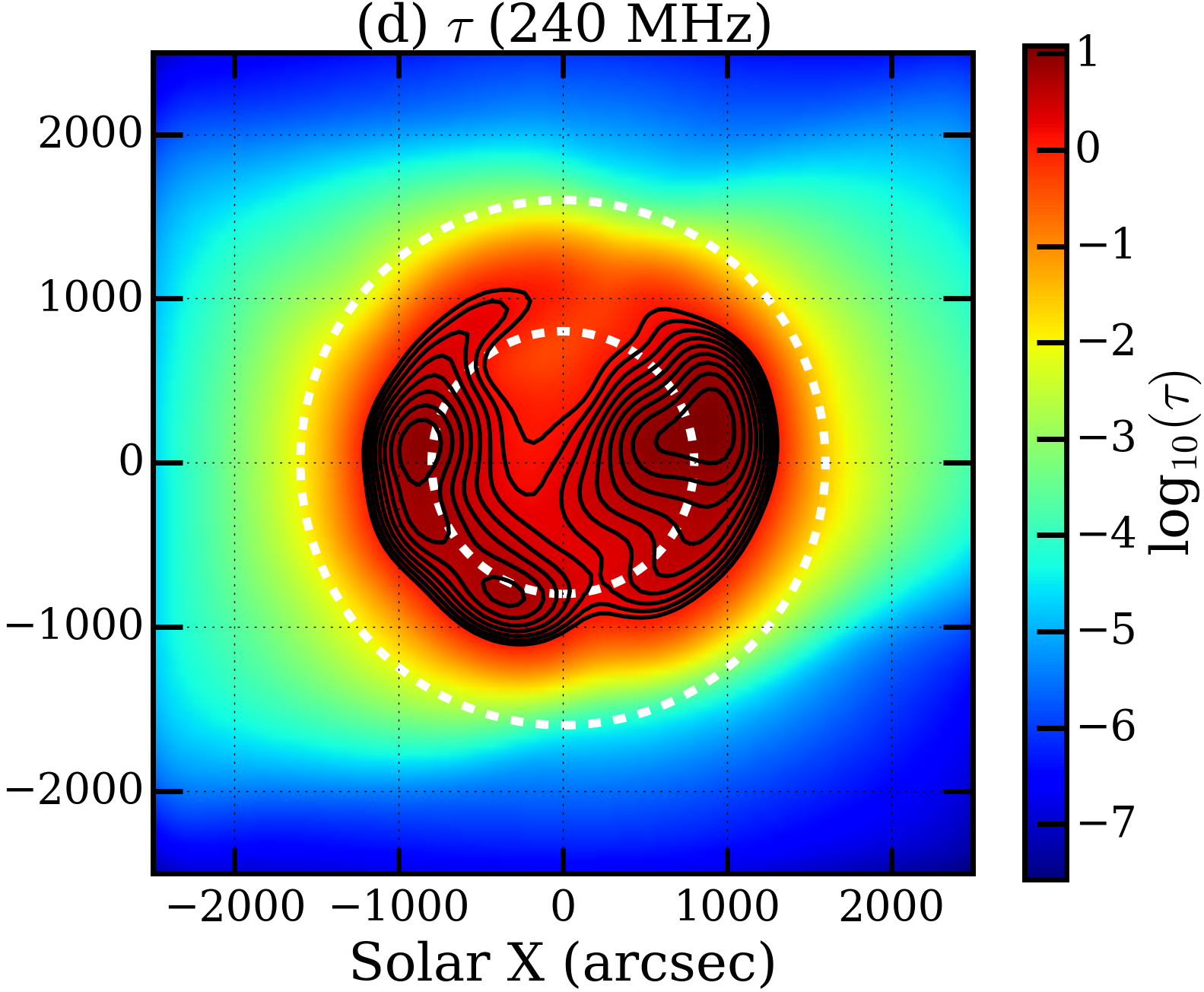}} 
\end{tabular}
\caption{From left to right, the panels show 
LoS averaged $n_e$, $T$ and $|B|$ maps obtained using PSIMAS.
The right most panel shows the $\tau_{ff}$ at 240 MHz obtained using FORWARD, convolved with the corresponding MWA PSF.
The white dotted circle marks the photospheric limb and the 10 contours are distributed uniformly between 10\% to 90\%.
}
\label{Fig:fwd_param}
\end{figure*}

Following the terminology introduced by \citet{Gibson2016} the coronal distribution of $n_e$, $T_e$ and $B$ can be said to define the ``physical state" of the system.
For the date of our observations, the physical state is determined using the MAS model for solar corona.
MAS self-consistently solves MHD equations in the 3D spherical coordinates of the solar corona/inner heliosphere\footnote{\texttt{https://www.predsci.com/corona/model\_desc.html}}.
As input, MAS uses the synoptic HMI magnetogram in Carrington coordinates.
The physical parameters $n_e$, $T_e$ and $B$ are self-consistently evolved globally using resistive thermodynamic MHD equations. 
The simulated values of $n_e$, $T_e$ and $B$ are normalized with photospheric values.  
Over the years, the results from this approach has been benchmarked with observations in various wavebands \citep[e.g.][etc.]{Lionello2009,Linker2011}. Once the physical state of the corona is available from this model, referred to as PSIMAS, the FORWARD package \citep{Gibson2016} is used to synthesize the radio images.
FORWARD is a comprehensive IDL package for coronal magnetometry which uses the physical state of the corona to synthesize a broad range of coronal observables, including radio images, using various ``physical processes".
The relevant physical processes for radio emissions are the thermal bremsstrahlung or the free-free emission and the gyroresonance emissions, though the contribution of the latter is significant only at much higher frequencies than considered here. The synthetic observables obtained, radio maps in this case, can be compared directly with the observations.

For thermal bremsstrahlung, which is always included in a FORWARD radio emission calculation, opacity results from collisions between electrons and ions.
The free-free emission optical depth ($\tau_{ff}$) originating from collisions of coronal ions and electrons \citep{Dulk1985,Gelfreikh2004}, which is used by the FORWARD package \citep{Gibson2016}, is given by
\begin{equation}
    \tau_{ff} = 0.2 \int \frac{n_{e}^{2}}{T_{e}^{3/2}(f\pm f_{B,LoS} )^{2}} dS,
    \label{Eq:tauff}
\end{equation}
where $S$, $f$ and $f_{B,LoS}$ are the coordinate along the LoS, the frequency of observation and the LoS gyro-frequency respectively in c.g.s units. 
The plus and minus signs represent the $o-$ and the $x-$modes respectively. 
Here $f_{B,LoS}=f_{B} |cos(\theta)|$, where $\theta$ is the angle between LoS and the local magnetic field $B$, and $f_{B} (Hz) = 2.8\times10^{6}\ B$, with $B$ expressed in Gauss.
FORWARD does not incorporate any propagation effects, i.e. it does not take in to account the scattering and refraction caused by the coronal density structures that are a part of the PSIMAS model or the refraction due to the large-scale density decrease with coronal height.
The $\tau_{ff}$ is computed simply using rectilinear propagation of radio waves through the coronal volume.
It also computes the $T_B$, the brightness temperature associated with a given ray, and provides 2D Plane-of-Sky (PoS) images of $T_B$ in various Stokes parameters.
The $n_e$, $T_e$, $\vec{B}$ vary across the LoS, and it is not straightforward to visualize these quantities in a 3D volume surrounding the Sun.
To provide a visual sense for the spatial variations of these parameters, Fig. \ref{Fig:fwd_param} shows the PoS maps of various LoS averaged quantities determined using PSIMAS and extracted using FORWARD. 
The first three of these, i.e. $n_e$, $T$ and $|B|$, define the physical state of the corona, and the last one, $\tau_{ff}$ at 240 MHz, is obtained by using the physical processes of thermal bremsstrahlung.
The $\tau_{ff}$ map has been convolved with the corresponding MWA point-spread-function (PSF) to facilitate comparison with MWA images in Section \ref{Sec:radio-analysis}.
A distinct circular feature is found to be associated with the solar limb in all of these  LoS integrated quantities.
As one proceeds from the centre of the sun towards the limb, the lengths of the LoS segment between any two coronal heights become increasingly longer.
This leads to a tendency for the $\tau_{ff}$ to increase with increasing distance from the centre.
As the LoS goes past the limb, its length through the corona abruptly increases, as the part of corona ``behind" the plane of the sky also becomes accessible to the LoS, giving rise to a pronounced limb ``brightening". 
Its smeared out appearance in the right most panel of Fig. \ref{Fig:fwd_param} is due to the convolution with the MWA PSF.
A comparison of Fig. \ref{Fig:fwd_param} with the Fig. \ref{Fig:aia_hmi} (right panel), shows that the coronal hole region is associated with low $n_e$, low $T$ and a somewhat higher $|B|$.
In the vicinity of the active regions on the western limb, an increase in $T$ and $|B|$ are evident.
Except for what has been noted above, the solar disk shows features with low contrast for $n_e$ and $|B|$, while the $T$ structure is more elaborate.
$\tau_{ff}$ map at 240 MHz clearly shows the expected limb brightening and a signature of the coronal holes. 
We divide the LoS into N segments with a given segment denoted by $i$, and $i=0$ denoting the base of the corona and $i=N-1$ the edge of the simulated volume closest to the observer.
The change in $T_B$ as the radiation traverses the $i^{th}$ segment of the LoS is given by 
\begin{equation}
    T_{B,i} = T_{B,i-1} e^{-d\tau_{i}} + T_{e,i} (1-e^{-d\tau_{i}})
    \label{Eq:Tb}
\end{equation}
where $d \tau_{i}$ is the $\tau$ corresponding to the $i^{th}$ segment of the LoS, and $T_{e,i}$ its electron temperature.
The PoS map is produced by integrating along the entire LoS using the above equation. Here we only use Stokes I $T_B$ maps at MWA frequency bands generated using FORWARD.

\section{Radio analysis}
\label{Sec:radio-analysis}

\begin{figure*}
    \centering
    \begin{tabular}{cc}
    \includegraphics[width=1.0\columnwidth]{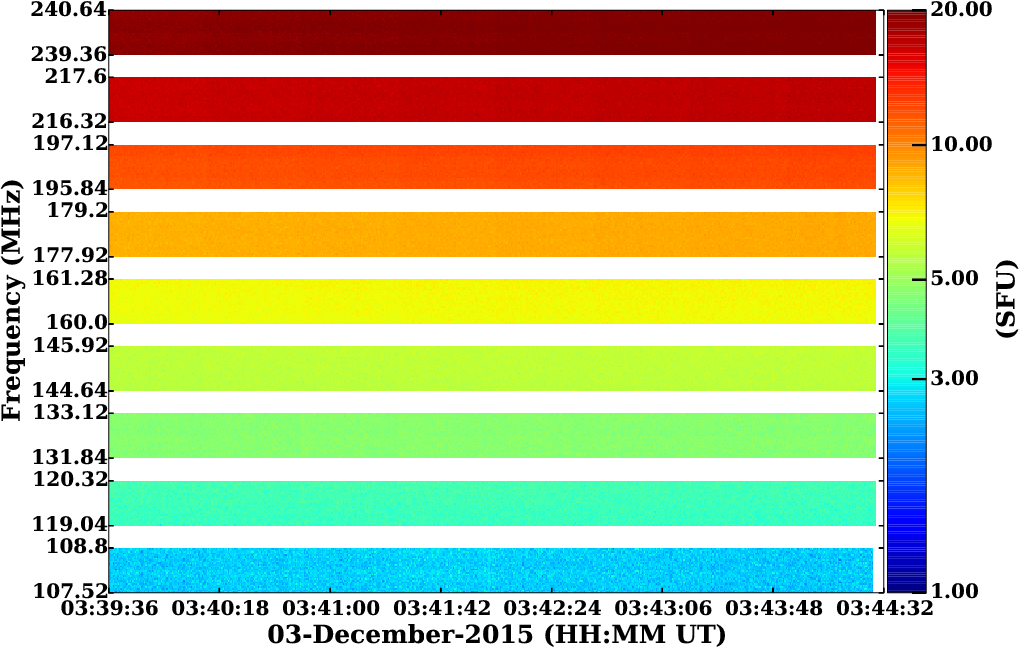}
         &  
 \includegraphics[width=0.7\columnwidth]{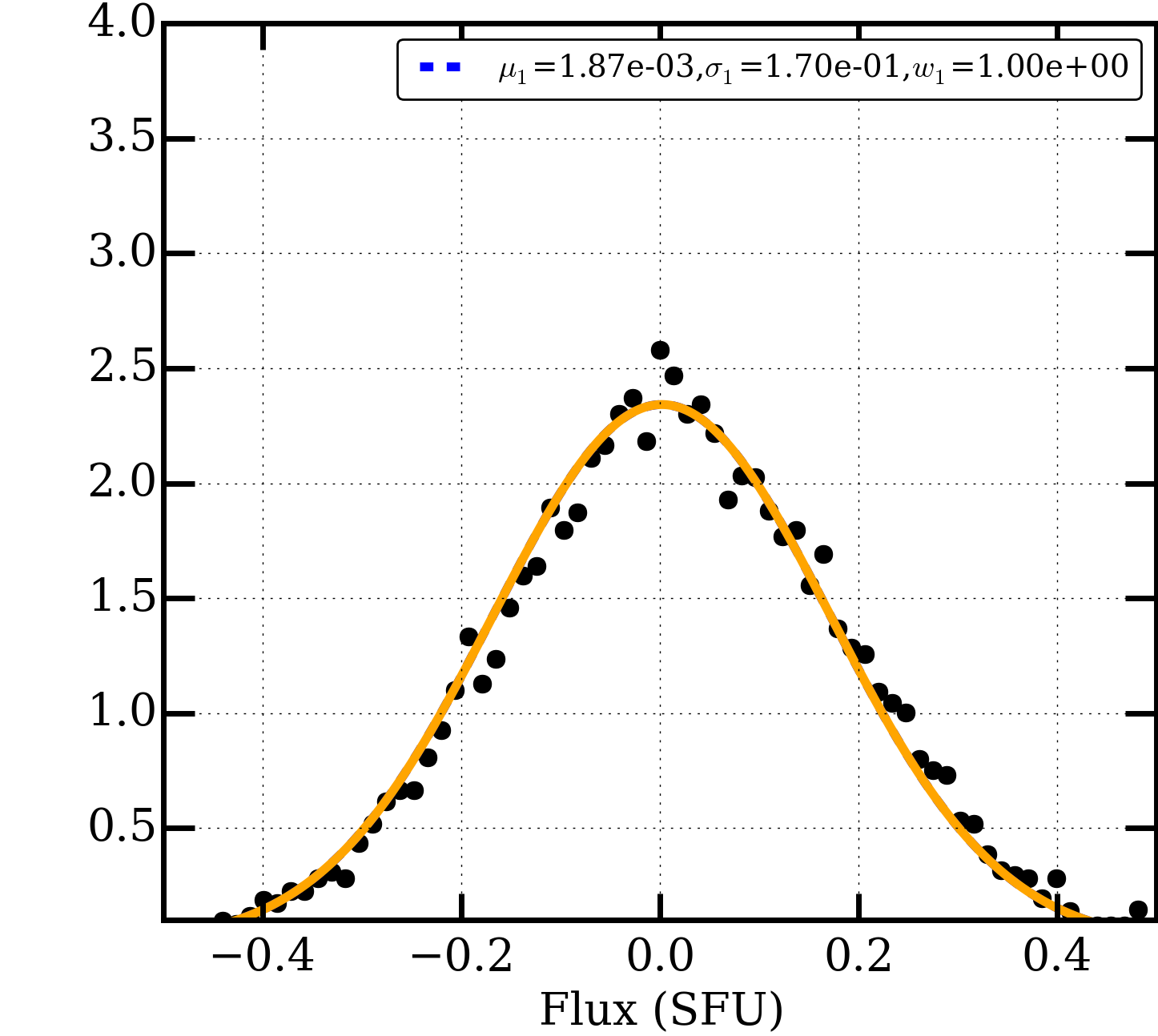}\\
     (a) Full Sun MWA DS & (b) 160 MHz   
    \end{tabular}
    \caption{Left panel: Flux calibrated MWA dynamic spectrum. 
    Right panel: Mean subtracted flux density distribution for an example band centred close to 160 MHz.
    The black points show the observed distribution, and the orange curve shows optimal Gaussian mixtures model arrived at to describe these data.
    The mean, $\mu$, and standard deviation, $\sigma$, of the Gaussian are mentioned.
    For all the 10 frequency bands, the model invariably comprises only one Gaussian component, corresponding to the thermal emission.
    }
    \label{Fig:mwa_ds}
\end{figure*}

\begin{figure*}
\begin{tabular}{ccc}
& I. MWA maps &\\
\resizebox{50mm}{!}{
\includegraphics[trim={0.0cm 0cm 0.0cm 0.0cm},clip,scale=0.33]{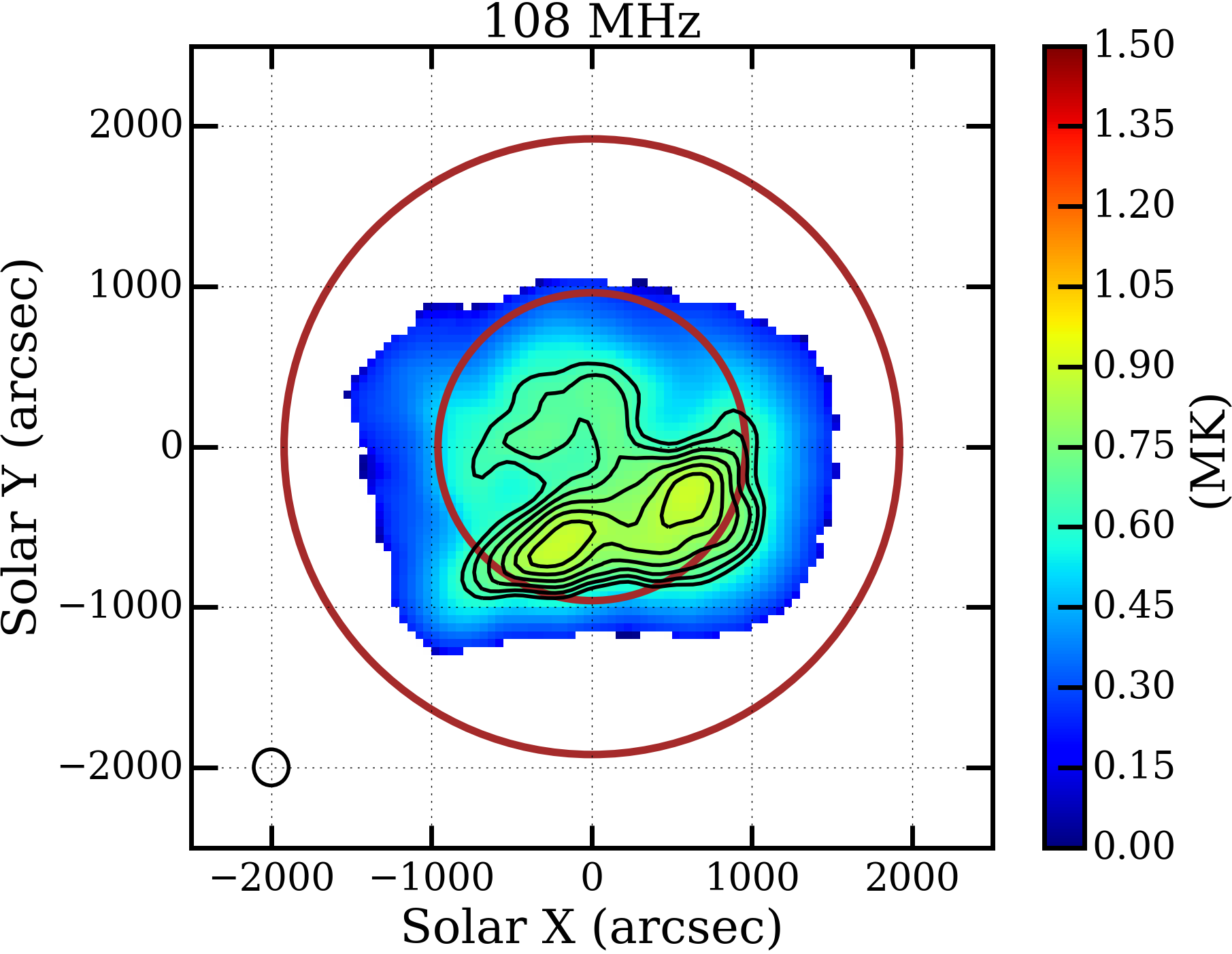}}
 &
\resizebox{50mm}{!}{
\includegraphics[trim={0.0cm 0cm 0.0cm 0.0cm},clip,scale=0.33]{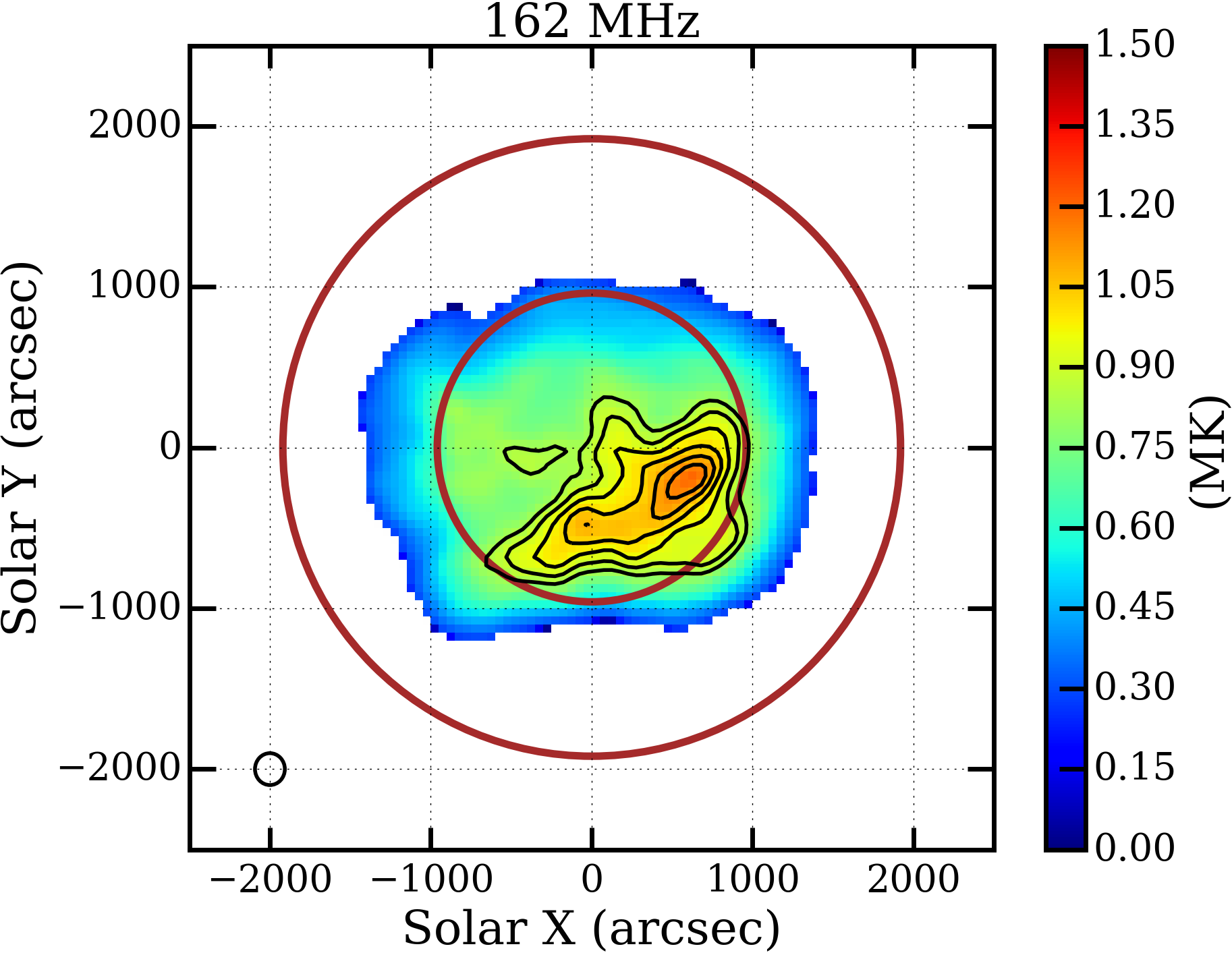}}
&
\resizebox{50mm}{!}{
\includegraphics[trim={0.0cm 0cm 0.0cm 0.0cm},clip,scale=0.33]{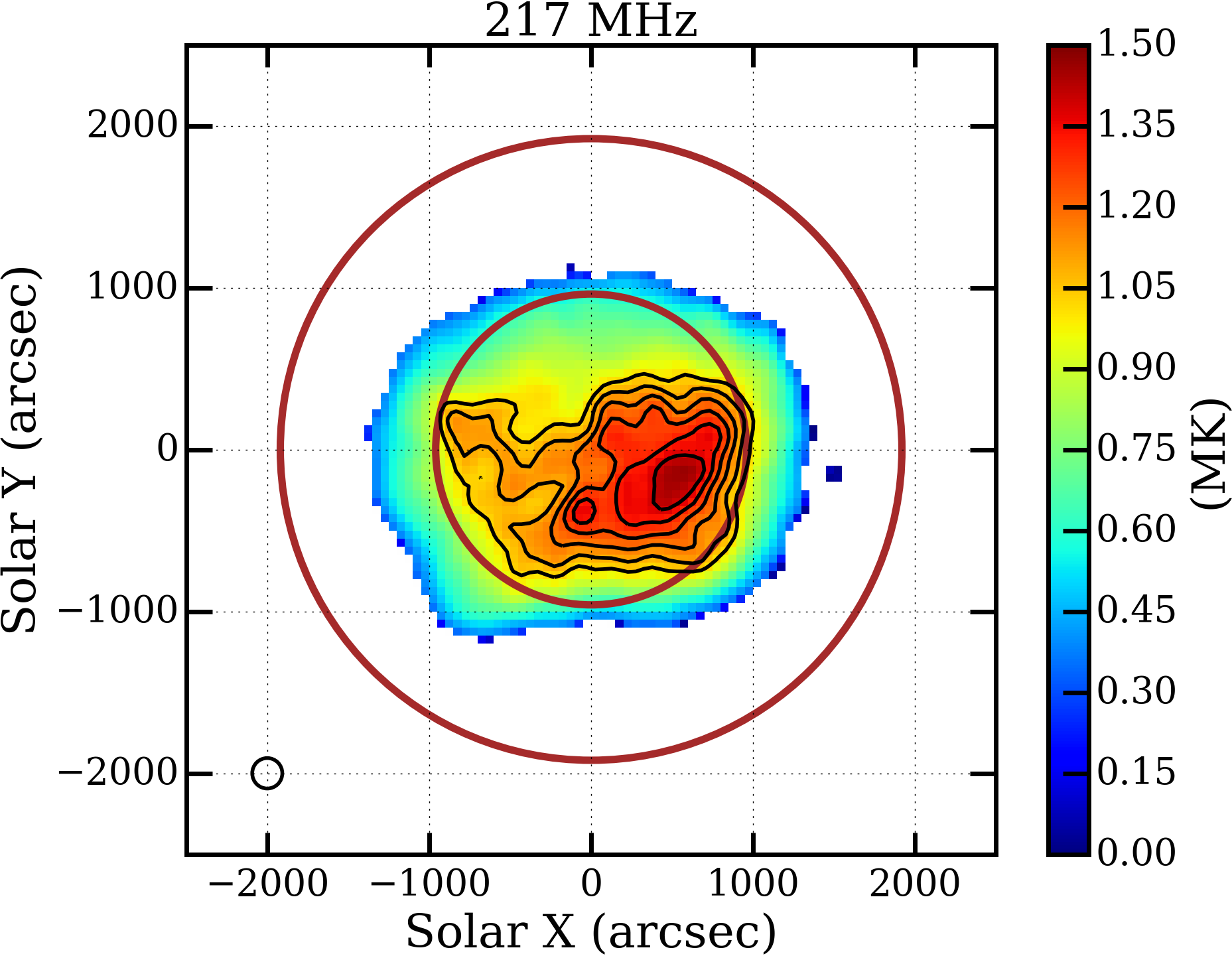}} \\
& II. FORWARD maps & \\
\resizebox{50mm}{!}{
\includegraphics[trim={0.0cm 0cm 0.0cm 0.0cm},clip,scale=0.33]{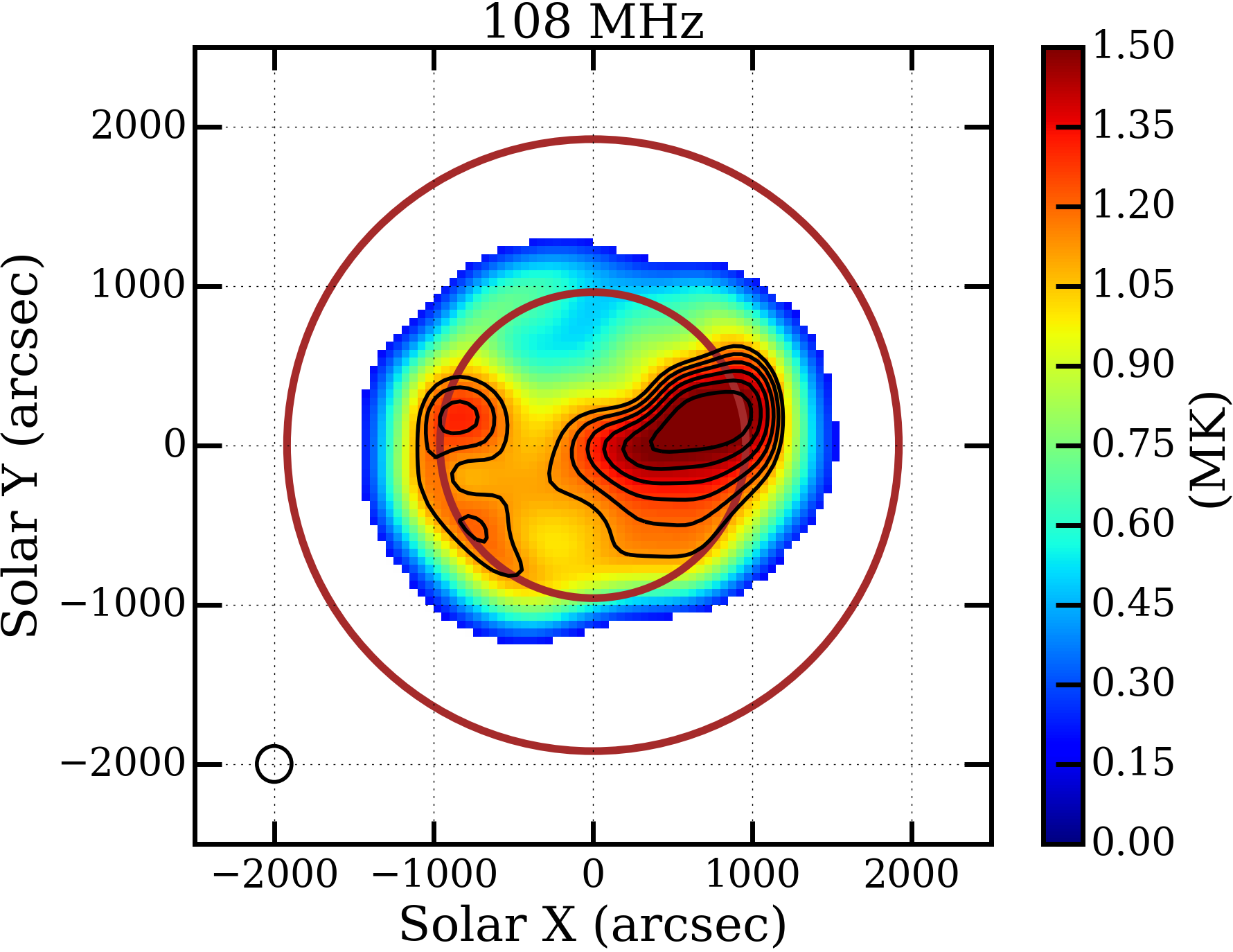}}
 &
\resizebox{50mm}{!}{
\includegraphics[trim={0.0cm 0cm 0.0cm 0.0cm},clip,scale=0.33]{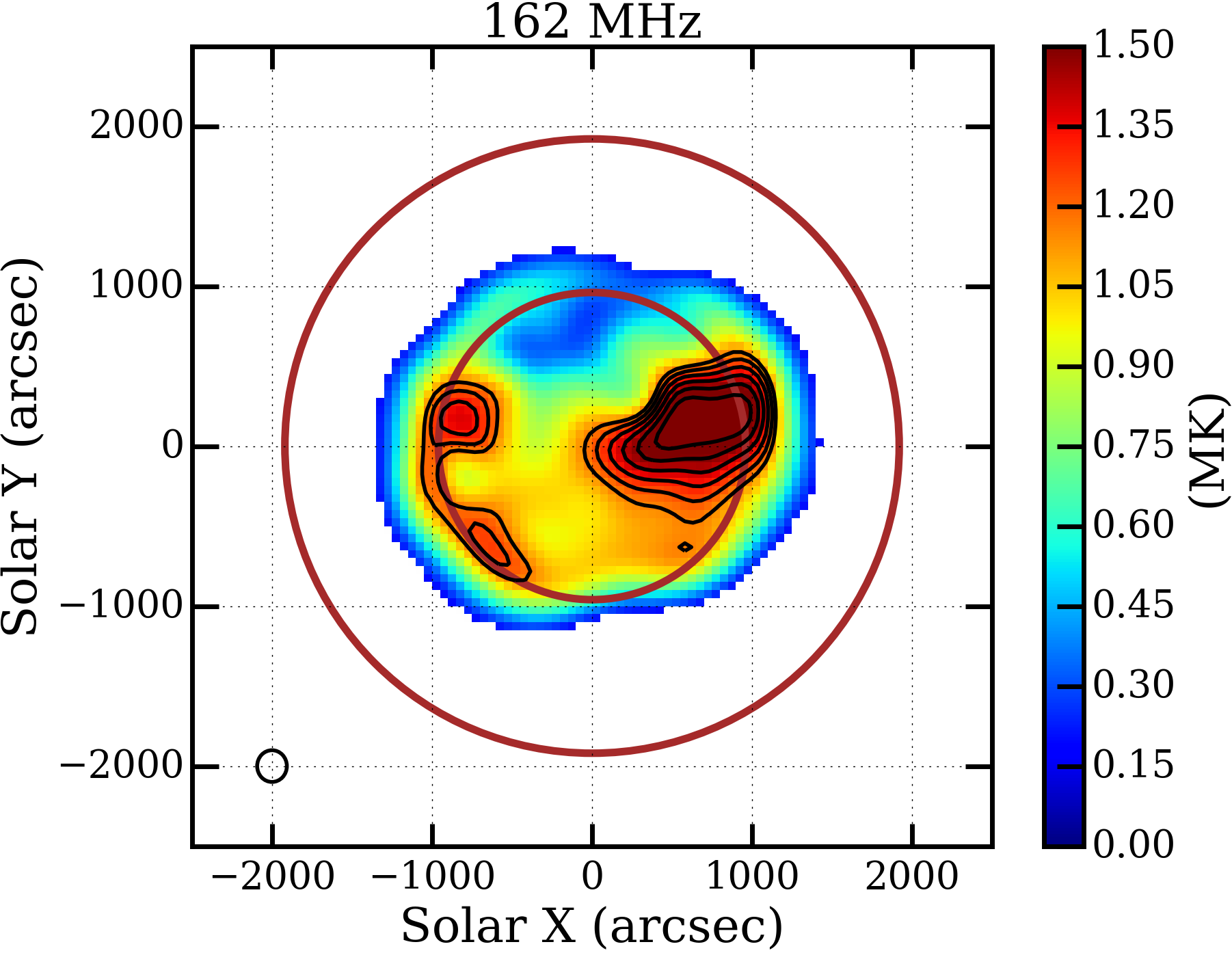}}
&
\resizebox{50mm}{!}{
\includegraphics[trim={0.0cm 0cm 0.0cm 0.0cm},clip,scale=0.33]{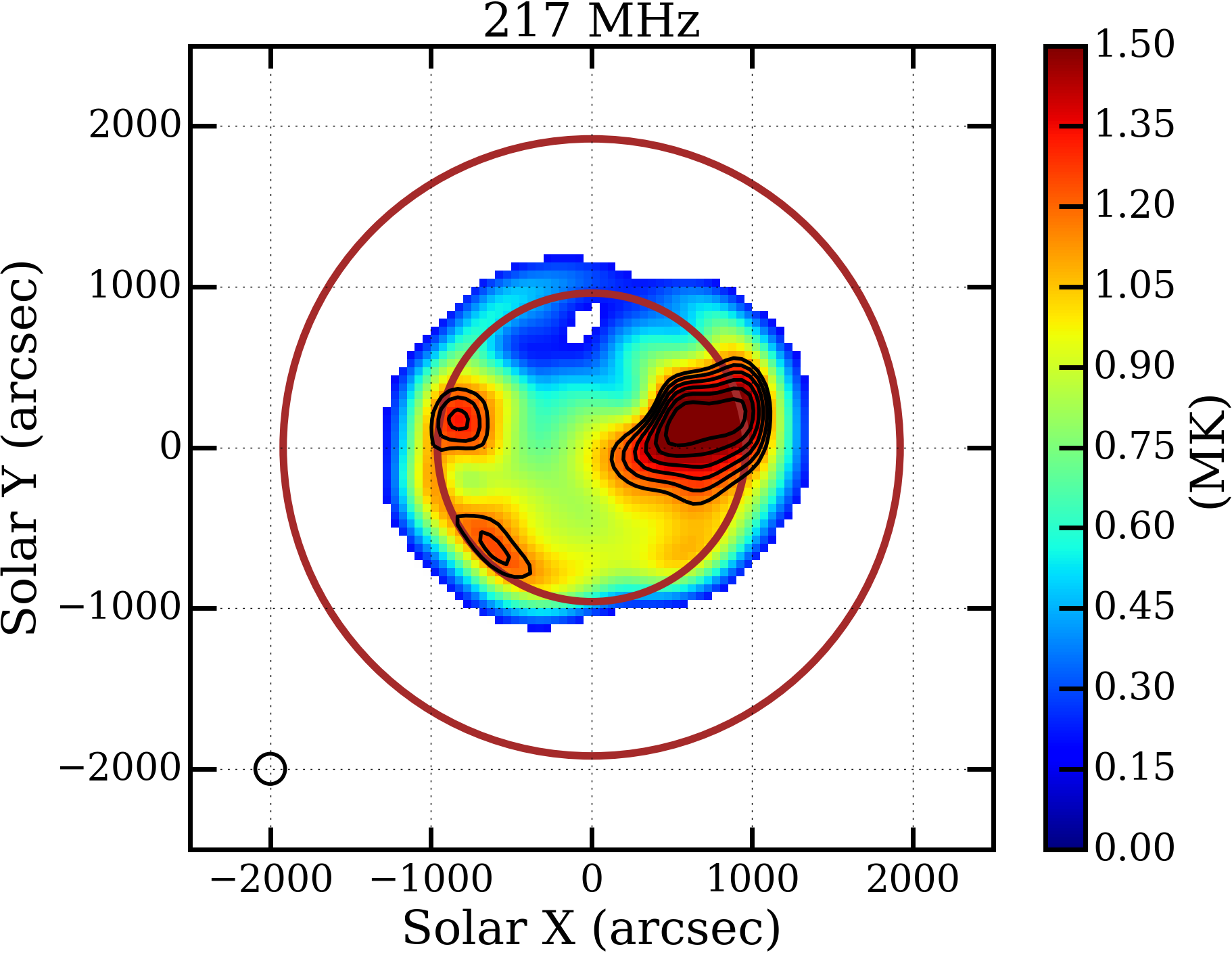}} \\
& III. EUV overlays & \\
\resizebox{45mm}{!}{
\includegraphics[trim={0.0cm 0cm 0.0cm 0.0cm},clip,scale=0.33]{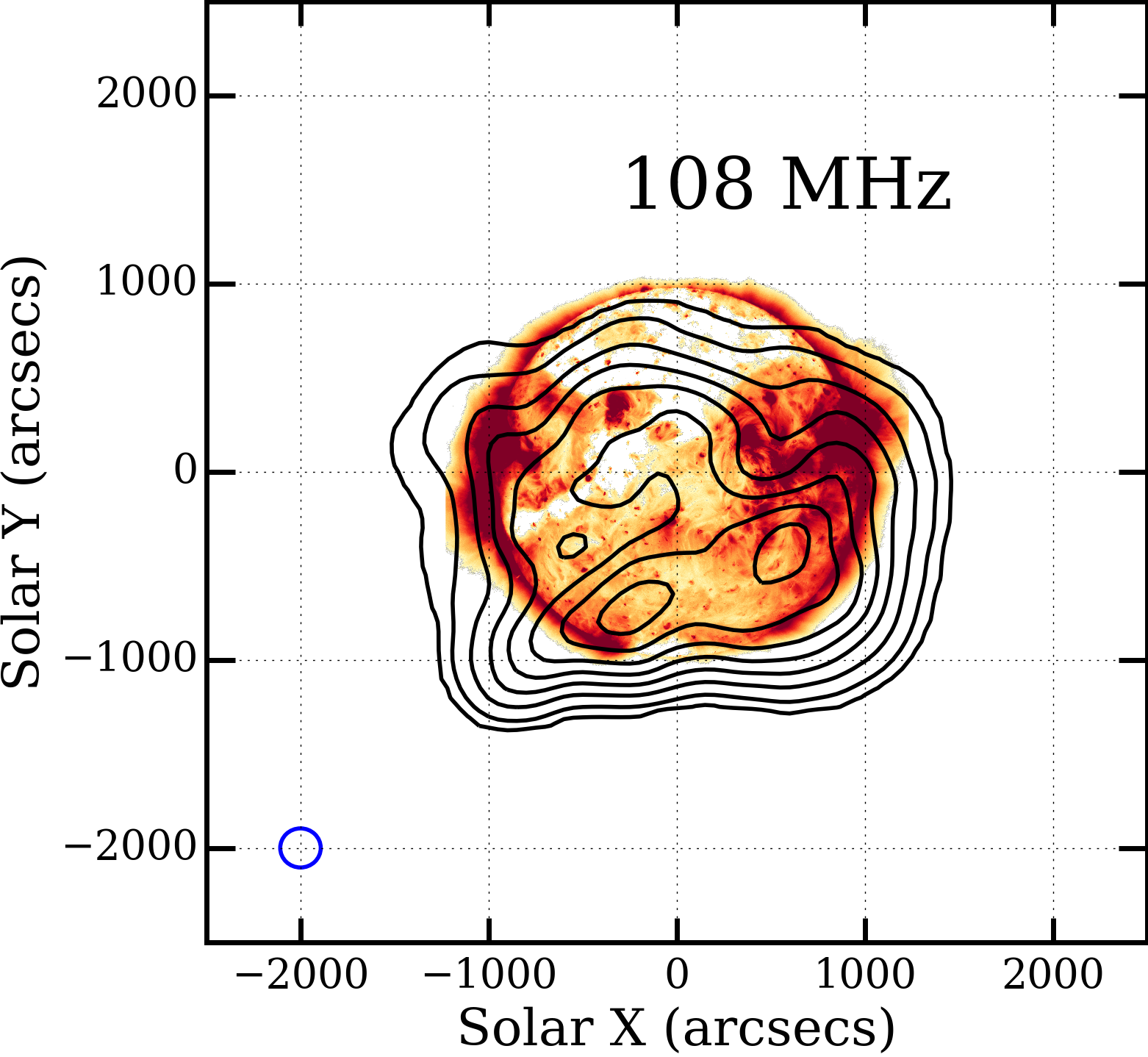}}
 &
\resizebox{45mm}{!}{
\includegraphics[trim={0.0cm 0cm 0.0cm 0.0cm},clip,scale=0.33]{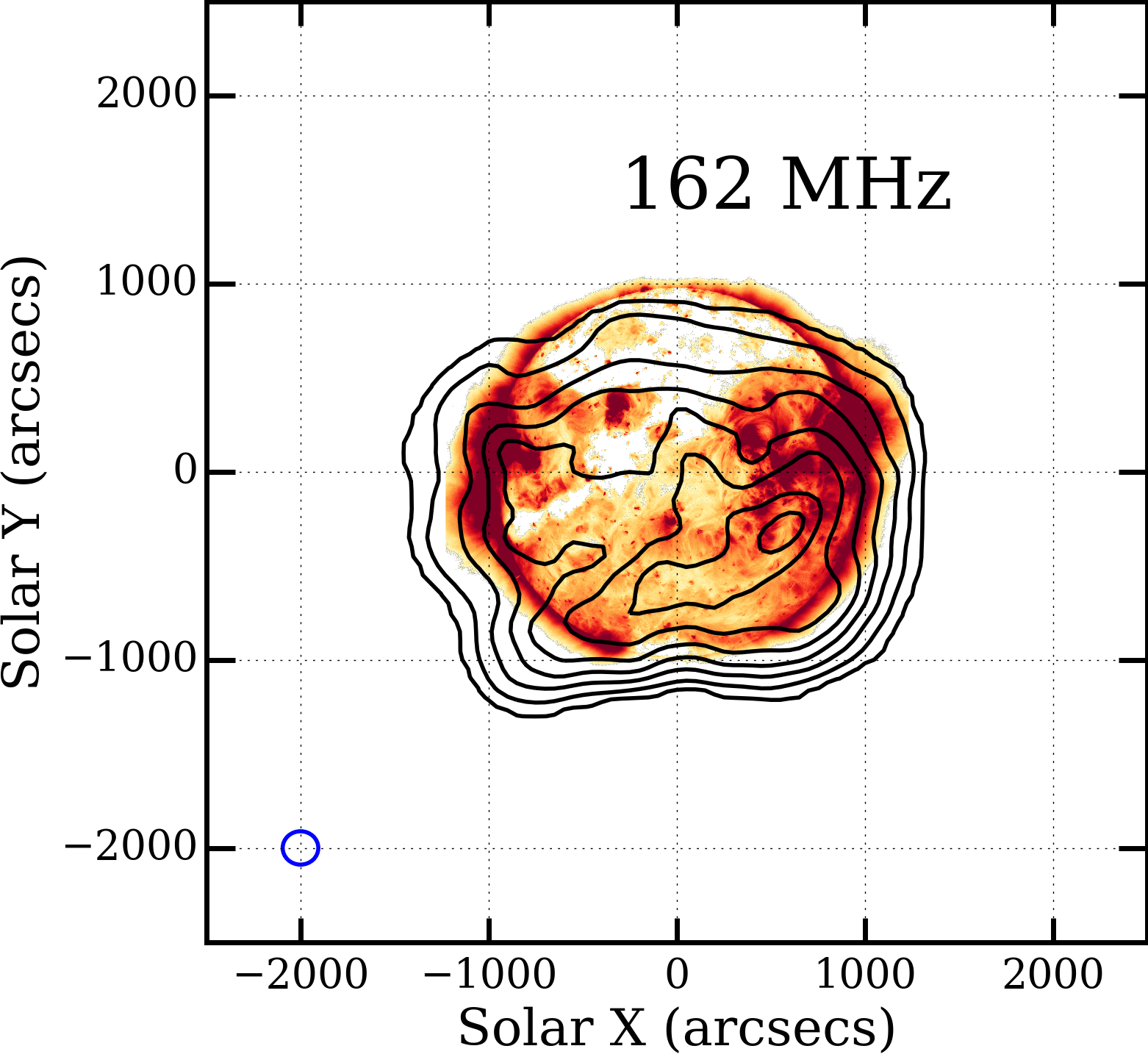}}
&
\resizebox{45mm}{!}{
\includegraphics[trim={0.0cm 0cm 0.0cm 0.0cm},clip,scale=0.33]{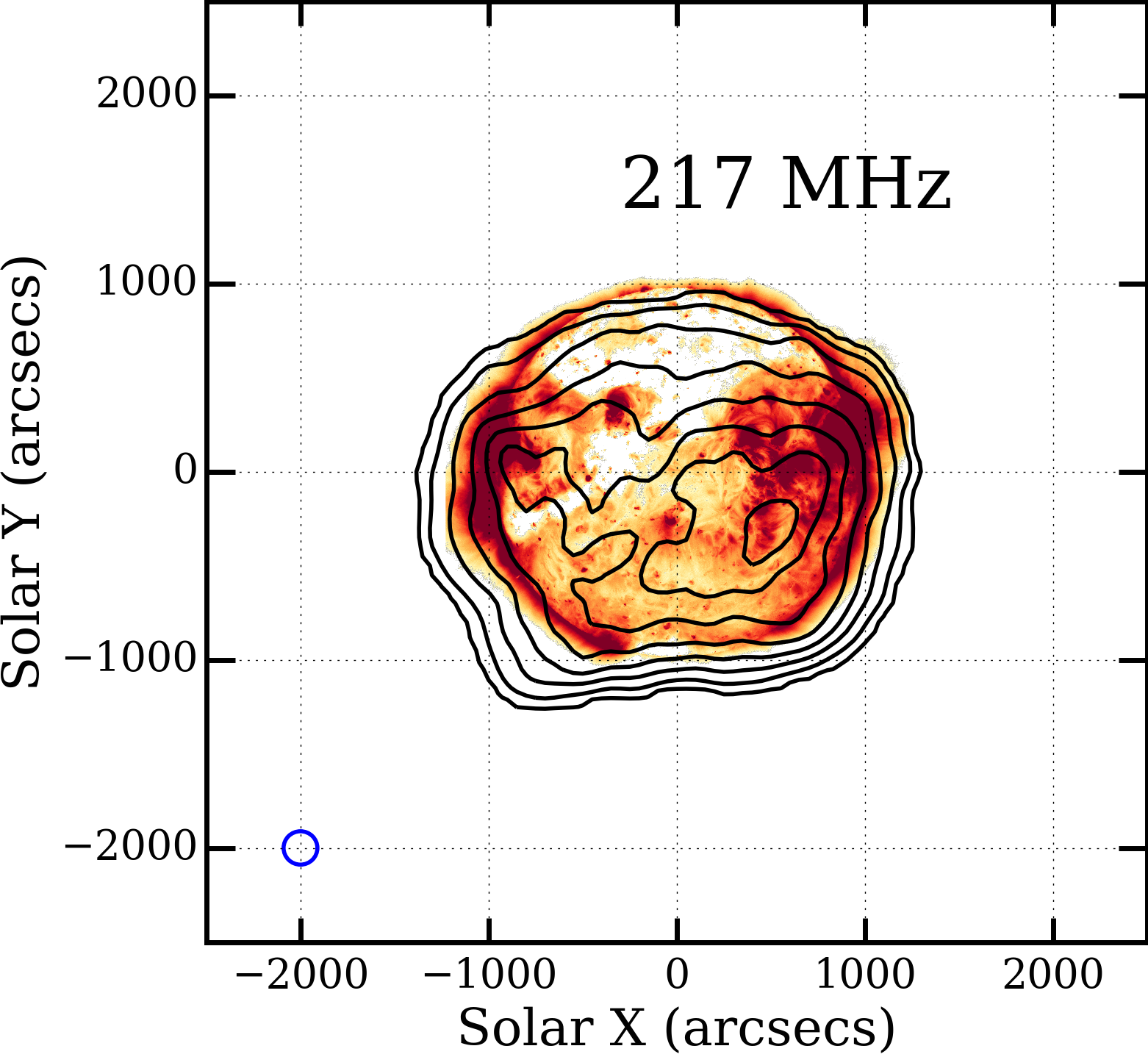}} 
\end{tabular}
\caption{Top panel: Averaged brightness temperature solar maps for three MWA frequency bands at 108, 162 and 217 MHz. The maps, which these are the average of, were made with frequency and time averaging of 2 MHz and 0.5 seconds, respectively. Middle panel: FORWARD brightness temperature solar maps for three frequency bands at 109, 162 and 217 MHz. Note that only  $T_B>0.2$ MK are shown. Circles mark 1 and 2 $R_{\odot}$. Contour levels in all the maps are at 70, 75, 80, 85, 90 and 95 \% of the peak. The peak MWA T$_{B}$ for 108, 162 and 217 MHz are 0.92, 1.20 and 1.50 MK, while for FORWARD maps peak T$_{B}$ are 1.59, 1.60 and 1.61 MK respectively. Bottom panel: MWA contours overlaid on AIA 193 \AA \ image for 108, 162 and 240 MHz. Contour levels are at 25, 35, 45, 55, 65, 75, 85 and 95\% of the peak $T_B$.
The bottom left ellipse in each panel show the corresponding PSFs.
\label{Fig:maps}}
\end{figure*}

\begin{figure}
\resizebox{70mm}{!}{
\includegraphics[trim={0.0cm 0cm 0.0cm 0.0cm},clip,scale=1.0]{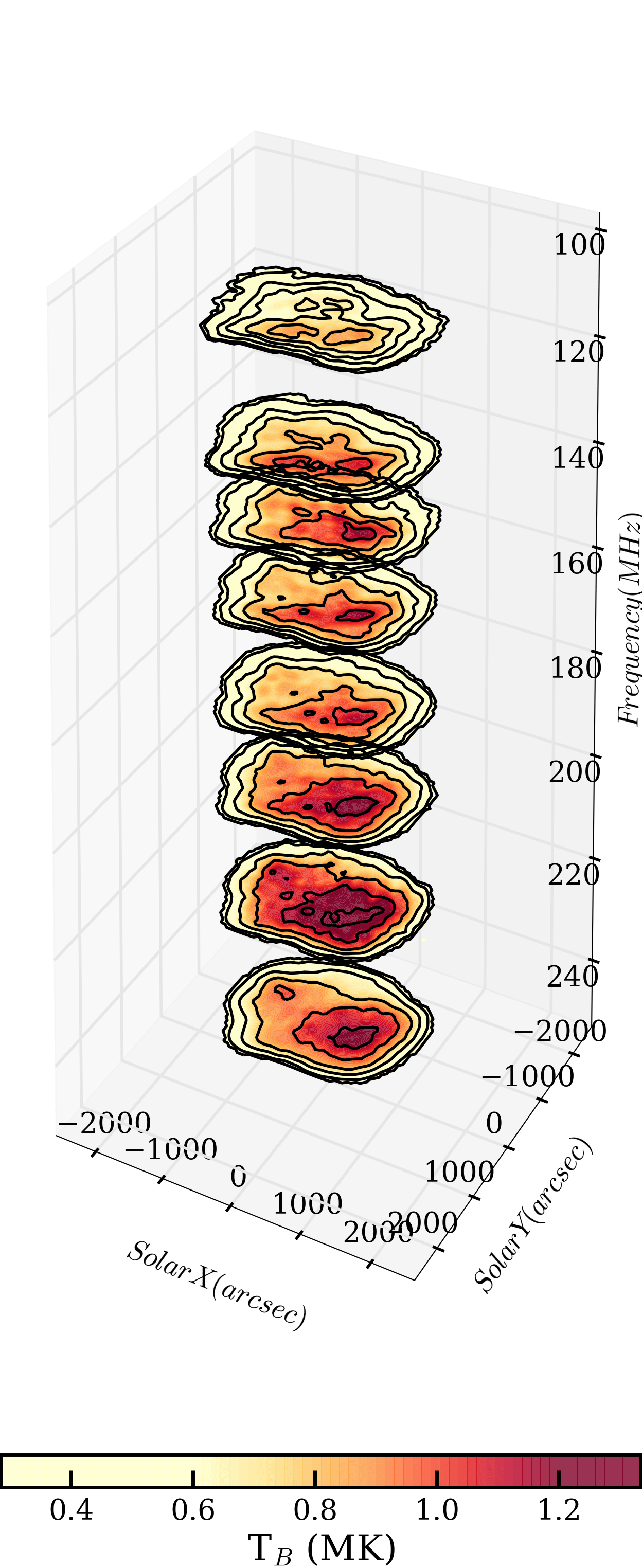}}
\caption{$T_B$ maps for 8 frequency bands from 109 MHz to 240 MHz. The frequency and time averaging are 2 MHz and 4 min respectively. The six contour levels corresponds to 10\%, 20\%, 40\%, 60\%, 80\% and 90\% w.r.t maximum $T_B$.
}
\label{Fig:Tb_all}
\end{figure}

We analyse $\sim$4 min of MWA data from  03:39:36 to 03:44:32 UT on 03 December, 2015.
The observations were done in the so called `picket-fence' mode, which allowed us to distribute the total available bandwidth of 30.72 MHz across the 80--240 MHz in 12 roughly log-spaced chunks, each 2.56 MHz wide.
Flux calibration was carried out following the prescription given by \citet{Oberoi2017} for all ten frequency bands above 100 MHz.

The flux calibrated dynamic spectrum (DS) is shown in Fig. \ref{Fig:mwa_ds}, left panel.
The DS is remarkably featureless, and no short-lived narrow-band weak nonthermal emission features, which are often observed even during a period of low to moderate solar activity, are visually evident.
Examples of DS which show such emission features during periods of low activity are available in \citet[][]{Oberoi2011, Oberoi2017, Suresh2017, Sharma2018}.
A technique based on Gaussian mixture decomposition developed by \citet{Sharma2018}, designed to quantify the presence of weak nonthermal emission features, was applied to these data. 
This technique, which has been demonstrated to reliably identify nonthermal emissions as weak as $\sim$0.2 SFU (Solar Flux Unit; 1 SFU = 10$^4$ Jy) with a prevalence as low as $\sim$1\%, was unable to find any evidence for the presence of nonthermal emissions (Fig. \ref{Fig:mwa_ds}, right panel). 
The lack of even a hint of presence of a nonthermal tail in the histograms for each of the ten frequencies analysed confirms that these data come from a period of exceptionally low solar activity, and that it is reasonable to regard the solar emission to be stationary over this 4 minute period.

Radio imaging was carried out using CASA \citep{McMullin2007} for eight frequency bands from 109 MHz to 240 MHz. We produced images, each with 2 MHz and 0.5 s bandwidth and time integration respectively. For each frequency band, we use different UV-ranges optimised for best images. These images were converted into $T_B$ maps following the procedure described by \citet{Mohan2017} for the 8 frequency bands. We used a 5$\sigma$ limit was to delimit the radio sun, where $\sigma$ is the RMS noise in the image far from the location of the Sun. Typical dynamic ranges of these images lies between 250--300.
Since focus of this work is the morphology of the $T_B$ maps, and there is little evidence for temporal variations, for every frequency, we work with the mean $T_B$ map obtained by averaging in time over all of the 0.5 s individual maps obtained during this 4 min duration.

\subsection{MWA brightness temperature maps}

The top row of Fig. \ref{Fig:maps} and Fig. \ref{Fig:Tb_all} show MWA $T_B$ maps.
Figure \ref{Fig:maps} shows radio maps for three representative observation frequencies - 108, 162 and 217 MHz.
Figure \ref{Fig:Tb_all} shows a stack of $T_B$ maps at all of the frequencies for which images were obtained, with the frequency of observation shown along the vertical axis. 
This figure takes advantage of the fact that the emission at higher radio frequencies is expected to arise at lower coronal heights to provide a convenient way to track the evolution of the radio morphology with coronal height. 
It is evident from these figures that the thermal bremsstrahlung from the disk dominates the emission.
A gradual and systematic evolution of the single dominant compact emission feature in the western hemisphere, which splits into two features at higher coronal heights, is clearly seen.
An increase in the size of the sun with height is also evident.
The 5$\sigma$ threshold used to delimit the Sun is $\approx$0.2 MK.
The solar size at the equator is larger than that at the poles, and this is more prominent at lower frequencies. 
An overall drop in $T_B$ of the compact source as well as extended emission features with increasing coronal height is also evident. 
However, the spatial $T_B$ distribution is rather smooth, with the contrast between the on-disk features even at the highest frequency being only about a factor of two, which steadily decreases to about 25\% by 108 MHz. 

\subsection{Limb Brightening}
\begin{figure}
    \centering
\resizebox{80mm}{!}{
\includegraphics[trim={0.0cm 0cm 0.0cm 0.0cm},clip,scale=0.33]{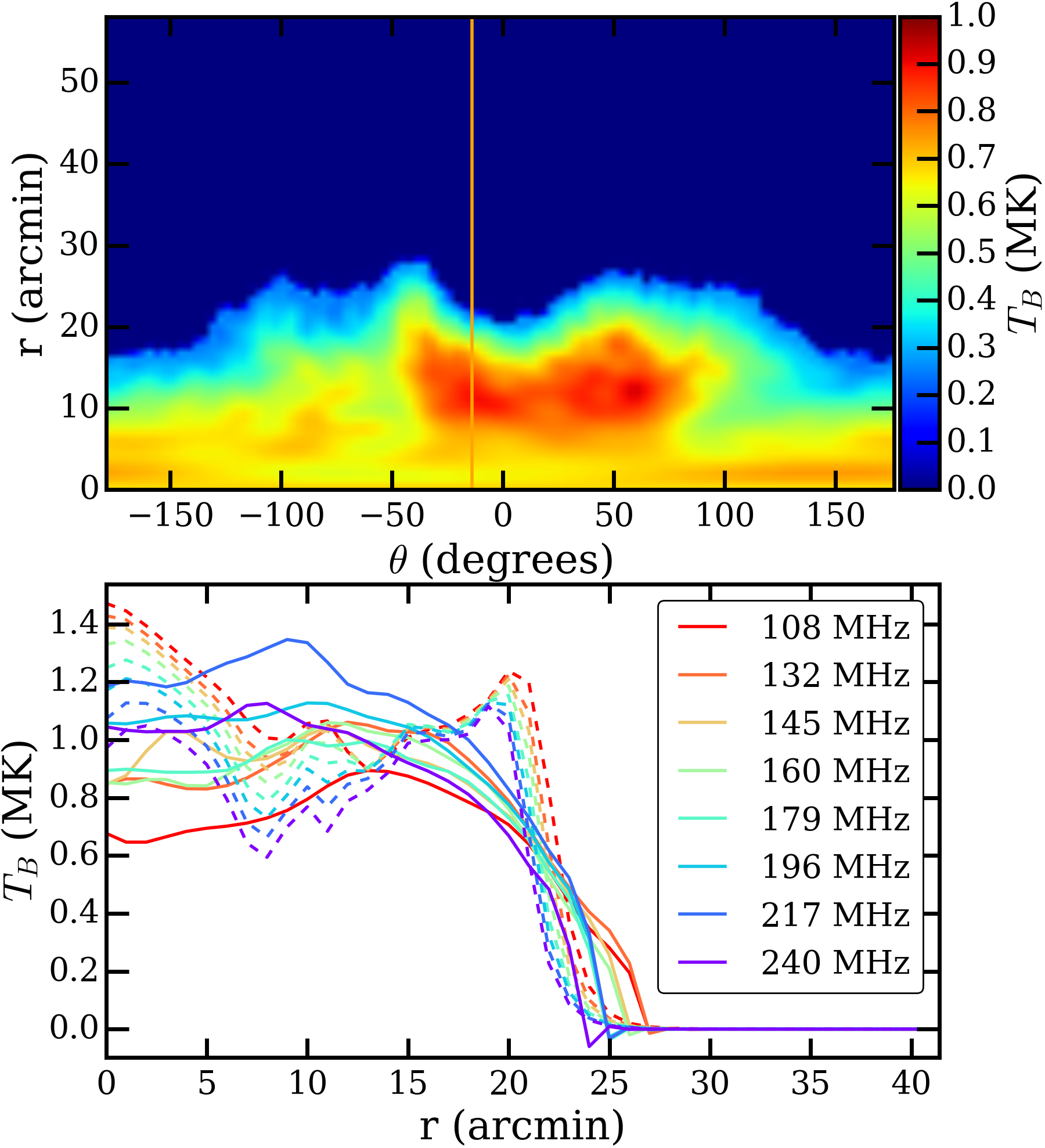}}
    \caption{108 MHz MWA map in radial and azimuthal coordinates. The south pole is at $\theta=0^o$. $\theta$ measures positive values in anti-clockwise direction. Bottom panel shows the cut along the radial coordinate at $\theta = -13.6^o$ for all MWA frequencies showing limb brightening. The solid and dashed lines corresponds to MWA and FORWARD maps respectively.
}
    \label{Fig:108MHz}
\end{figure}

Limb brightening is clearly evident in the FORWARD $\tau_{ff}$ map  (Fig. \ref{Fig:fwd_param}, right most panel, eastern limb), though it is less prominent in the FORWARD radio maps (Fig. \ref{Fig:maps}, middle row).
As pointed out in Sec. \ref{subsec:fwd_model}, the ray paths used by FORWARD for computing the observed maps do not take propagation effects into account and are hence unrealistic.
However, detailed studies taking coronal refraction into account also show limb brightening at frequencies ranging from 1200 MHz to below 200 MHz \citep{Smerd1950}.
As an impact of the large scale refraction, the location of the peak of the brightness profile is seen to shift closer to the solar centre with decreasing frequency.
It is quite pronounced at decimeter wavelengths and has been observed \citep[e.g.][]{Swarup1955}.
At meter wavelengths however, the limb brightening is much less pronounced, and to the best of our knowledge, is yet to be reported observationally.
In addition to scattering, which is expected to smear out the limb brightening, there are also other factors which have lead to this.
These include the intrinsic difficulties in imaging the quiet sun with sufficient dynamic range, and the presence of contaminating coronal features like coronal holes to active regions.


To look for evidence for limb brightening, we examine a position angle far from contaminating effects of coronal features.
The top panel of Fig. \ref{Fig:108MHz} shows the 108 MHz $T_B$ map in radial-azimuthal coordinates.
The bottom panel shows the radial profiles at various frequencies along the azimuth of $\theta = -13.6^{\circ}$ (lying in the south-eastern quadrant) for both, MWA and FORWARD maps. 
The FORWARD maps show clear presence of limb brightening between 20' and 25'.
Similar features are not seen in the MWA maps.
Much weaker $T_B$ enhancements are seen between 10' and 15' in the MWA maps.
These could be the signatures of residual limb brightening much reduced by propagation effects.
We note the tendency for the peak of the $T_B$ profile to move to lower radii with increasing frequency. 
Some other expected tendencies, like the increase in radial size and the decrease in observed $T_B$ with decreasing frequency are also evident.

\subsection{Radio-EUV overlays}

Bottom row of Fig. \ref{Fig:maps} shows the overlays of radio $T_B$ contours over AIA EUV 193 \AA images. 
At the higher frequency end of our observations, the large scale correspondence between the EUV and radio emissions is clearly evident.
The EUV bright regions tend to also have regions of higher $T_B$ nearby.
The large coronal hole, occupying much of the northern hemisphere, is clearly seen in the EUV with a thin tongue dipping into the southern hemisphere. 
At 240 MHz, the coronal hole region is associated with lower $T_B$, though by 108 MHz, the same region has $T_B$ higher than its neighbouring regions.
In fact, at 108 MHz the enhanced $T_B$ sits right atop the thin coronal hole tongue.

This tendency for some coronal holes to transition from regions of relatively lower $T_B$ to regions of relatively higher $T_B$ as one goes from lower to higher coronal heights has been noticed earlier as well \citep[e.g.][]{Mercier2015} , and was the subject of a recent study by \citet{Rahman2019}, who proposed an explanation based on refractive effects.

\subsection{Effect of refraction and scattering}

The significant impact of propagation effects has long been recognised for the active emissions arising from plasma emission mechanisms, where the generated radiation is at the local plasma frequency or its harmonic \citep{Steinberg1971,Arzner1999,Thejappa2008,Kontar2019}.
These propagation effects are equally applicable to the quiet emissions as well

The top and middle rows of Fig. \ref{Fig:maps} show the $T_B$ maps observed by MWA and those computed by FORWARD, respectively.
The FORWARD maps have been convolved with the MWA restoring beam to facilitate comparison.
At 217 MHz, it seems reasonable that the impacts of scattering and refraction can morph the FORWARD map to look like what is observed by the MWA.
The FORWARD maps show a larger contrast, and comparatively compact emission features at the locations of active regions present close to the limb, as seen in the EUV maps.
The large scale refraction would tend to move features at the limb to smaller radial distances, and the scattering would smear them over a larger angular size.
As one proceeds to lower frequencies, the variance between the MWA and FORWARD maps grows.
While the FORWARD maps are all very similar. 
In the MWA  maps, with decreasing frequency the emission from the active region on the north-eastern limb disappears, a new compact source appears in the south-eastern quadrant, and the coronal hole transitions from being radio faint to be being radio bright.
While the peak $T_B$ of FORWARD maps is close to 1.6 MK at all frequencies, for the MWA it drops from 1.5 MK to 1.2 and 0.9 MK as one goes from 217, to 162 and 108 MHz respectively. 
The disk-averaged $T_B$ for FORWARD, $T_{B,FWD}$ and MWA, $T_{B,MWA}$ maps are listed in Table \ref{Tab:temp}.
$T_{B,MWA}$ is consistently significantly less than $T_{B,FWD}$, except at the highest two frequencies, where the two match within the uncertainties.

\begin{table*}
\caption{The Brightness temperature calculated from MWA and FORWARD, along with their optical depth estimates. 
The $\tau_{MWA}$ is calculated using Equation \ref{Eq:tauff}, while $\tau_{FWD}$ is the free-free optical depth obtained from the FORWARD code. 
$\Phi_{MWA}$ and $\Phi_{FWD}$ are the areas of the radio solar disk calculated from the MWA and FORWARD images respectively.
r$_{eff, MWA}$ is the effective radius corresponding to $\Phi_{MWA}$. 
Note: We use 5-$\sigma$ cut-off to define radio Sun, where $\sigma$ is the RMS computed over a region of the map far from the Sun. 
The last column lists the ratio of disk integrated flux densities from FORWARD and MWA maps, $\gamma = S_{\nu, FWD}/S_{\nu, MWA}$}
\begin{center}
\begin{tabular}{|c|c|c|c|c|c|c|c|c|c|c|c|}
\hline
Frequency & T$_{B,MWA}$ & T$_{B,FWD}$ & $\tau_{MWA}$ & $\tau_{FWD}$ & $\Phi_{MWA}$ & r$_{eff, MWA}$  &  $\Phi_{FWD}$ & $\gamma$ \\
(MHz)  &  (MK)  &  (MK)  &  &  & ($10^3$ arcmin$^2$)  & (arcmin) & ($10^3$ arcmin$^2$) & \\
\hline
108  &  0.51 $\pm$ 0.06  &  0.86  &  0.901 $\pm$ 0.005  &  13.46  &  2.33 $\pm$ 0.05  &  27.20 $\pm$ 0.10  &  1.86 & 2.03 \\
132  &  0.62 $\pm$ 0.13  &  0.86  &  1.279 $\pm$ 0.005  &  9.48  &  2.26 $\pm$ 0.05  &  26.85 $\pm$ 0.10  &  1.81 & 1.57 \\
145  &  0.63 $\pm$ 0.04  &  0.85  &  1.369 $\pm$ 0.011  &  7.98  &  2.24 $\pm$ 0.05  &  26.71 $\pm$ 0.09  &  1.78 & 1.52 \\
162  &  0.66 $\pm$ 0.05  &  0.85  &  1.471 $\pm$ 0.009  &  6.79  &  2.1 $\pm$ 0.05  &  25.87 $\pm$ 0.09  &  1.72 & 1.50 \\
179  &  0.65 $\pm$ 0.06  &  0.83  &  1.558 $\pm$ 0.009  &  5.54  &  2.11 $\pm$ 0.05  &  25.93 $\pm$ 0.09  &  1.63 & 1.44 \\
196  &  0.75 $\pm$ 0.07  &  0.81  &  2.625 $\pm$ 0.036  &  4.72  &  2.07 $\pm$ 0.05  &  25.65 $\pm$ 0.09  &  1.60 & 1.21 \\
217  &  0.86 $\pm$ 0.10  &  0.78  & $>1$  &  3.94  &  2.01 $\pm$ 0.04  &  25.29 $\pm$ 0.09  &  1.60 & 1.03 \\
240  &  0.75 $\pm$ 0.10  &  0.72  &  $>1$  &  3.15  &  1.97 $\pm$ 0.04  &  25.02 $\pm$ 0.09  &  1.51 & 1.02 \\
\hline
\end{tabular}
\end{center}
\label{Tab:temp}
\end{table*}

\subsubsection{Source sizes}
 A comparison of the observed and predicted angular sizes of the sun, can be a potentially useful way to quantify the impact of scattering.
Table \ref{Tab:temp} lists the solar disk area estimated from the MWA, $\Phi_{MWA}$, and FORWARD, $\Phi_{FWD}$, maps.
$\Phi_{MWA} > \Phi_{FWD}$ at all frequencies by $25-30\%$, with no significant trend with frequency.
For the MWA maps, a $5\sigma$ cutoff is used to define the extent of the radio Sun, where $\sigma$ is the RMS in the map in a region far away from the Sun.
For the FORWARD map, the radio sun is delimited using a $T_B$ contour corresponding to the $5\sigma$ value used for the MWA maps.

\begin{table}[]
    \centering
    \begin{tabular}{|c|c|c|c|c|c|}
 \hline
$\nu$  &  FORWARD  &  MWA  &  Ratio  &  Shift & Type III\\
 (MHz)   &   (PSFs)   &   (PSFs)   &  &  (arcmin) & (PSFs) \\
 \hline
108  &  2.65  &  5.5  &  2.08  &  10.44 $\pm$ 3.63 & 1.91\\
132  &  2.44  &  4.13  &  1.69  &  10.17 $\pm$ 3.27 & 1.72\\
145  &  2.34  &  3.92  &  1.67  &  8.58 $\pm$ 3.28 & 1.61\\
162  &  2.55  &  4.28  &  1.68  &  8.58 $\pm$ 3.09 & 1.48\\
179  &  2.29  &  4.02  &  1.76  &  10.87 $\pm$ 3.09 & 1.36\\
196  &  2.04  &  3.06  &  1.50  &  10.44 $\pm$ 3.11 & 1.27\\
217  &  1.83  &  2.39  &  1.31  &  8.33 $\pm$ 3.10 & 1.15\\
240  &  1.73  &  2.19  &  1.26  &  8.86 $\pm$ 3.37 & 0.94\\
 \hline
    \end{tabular}
    \caption{Source sizes of regions emitting equal flux density in FORWARD and MWA maps, as shown in Figure \ref{Fig:size}. 
    Note that the source sizes are given in units of the area of the PSF. 
    The ratio is simply the size seen by MWA divided by that estimated from FORWARD maps. The second last column shows the  distance between the locations of the peaks in FORWARD and MWA maps. 
    The size of the major axes of the PSFs are listed as the errorbars. 
    The last column shows the expected FWHM of the type-III burst sources in units of MWA PSFs for comparison. }
    \label{Tab:size}
\end{table}

\begin{figure*}
\begin{center}
\begin{tabular}{c}
\resizebox{120mm}{!}{
\includegraphics[trim={0.0cm 0cm 0.0cm 0.0cm},scale=0.8]{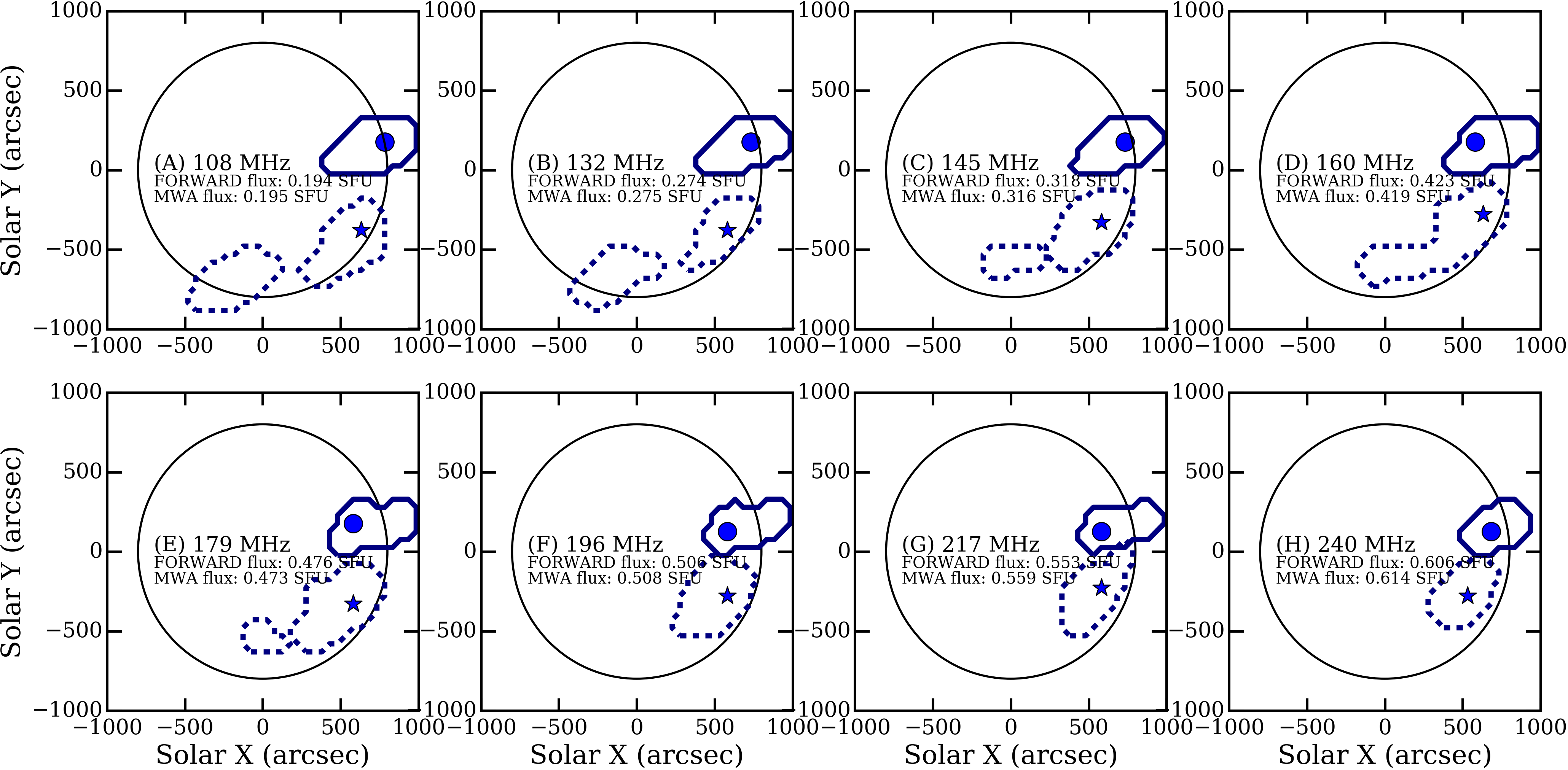}}
\end{tabular}
\end{center}
\caption{MWA and FORWARD contour maps for different frequencies. FORWARD contours are drawn at T$_{B}$=1.5 MK and shown by a solid line. 
The MWA contours, shown by dashed lines, have been drawn to enclose the same flux density as is enclosed in the FORWARD contour.
The filled circle and the star mark the locations of the flux density peak in the FORWARD and MWA maps respectively.
}
\label{Fig:size}
\end{figure*}

We also explore the use of the observed compact source to estimate the impact of scattering.
For this we define a region of size comparable to the PSF around the brightest source seen in the FORWARD maps, and then estimate the size of the corresponding region in the MWA maps which encloses the same radio flux density.
Though simple minded and really valid only for an isolated or a very bright compact source, it can still provide an interesting information about the impact of propagation effects.
Figure \ref{Fig:size} shows the regions in FORWARD maps marked by a contour at 1.5 MK for frequency bands by solid lines.
The area of this region varies between 1.73--2.65 times the PSF area for 240 MHz to 108 MHz (Table \ref{Tab:size}).
The regions marked by dashed contours are the regions from MWA maps around the peaks and enclosing the same flux density as the FORWARD contour.
The locations of the flux density peaks are also indicated.
The area enclosed by the MWA contours is always larger that enclosed by the FORWARD contours, by a factor between 1.26--2.08, and is listed in the column titled `Ratio' in Table \ref{Tab:size}. We note that as the FORWARD maps are convolved with MWA PSF, the true size of the source can be smaller and these numbers hence represent a lower limit.
At higher frequencies, the corresponding region in the MWA maps is a single unbroken contour. 
As one proceeds to frequencies below 196 MHz, another contour emerges in the MWA maps. 
In all likelihood, this is an unrelated emission component, and only illustrates the  the limitations of this exercise.
While the FORWARD maps are very similar across frequencies in their broad features, the MWA maps are not and some of these differences arise due to reasons other than propagation effects.

The impact of scattering on the type-III source sizes \citep{Kontar2019} can be modelled as,
\begin{equation}
    \Phi (^{\circ}) = (11.78\pm0.06) \times \nu^{(-0.98\pm0.05)}.
    \label{Eq:typeIII-size}
\end{equation} 
where $\Phi$ and $\nu$ are in degrees and MHz respectively.
Type III bursts are believed to arise from compact sources, so it is useful to use this as a point of comparison. 
The last column of Table \ref{Tab:size} lists the expected sizes of type-III sources based on Eq. \ref{Eq:typeIII-size} in units of the MWA PSFs for the corresponding frequency. 
The expected linear size of type III source is found to increase by $\sim$39\% from 240 to 196 MHz 
while an increase of $\sim$35\% in observed in the linear size of the compact source seen in MWA maps. 
The observed fractional change in the MWA source size is consistent with the observed type-III source size relation.

Another interesting finding from this exercise is that the location of the peak of the emission has substantial offsets between the two maps.
These offsets are tabulated in Table \ref{Tab:size} and lie between $\sim$8'--11', much larger than the $\sim$3' PSF.
The displacement vector has essentially the same orientation at all frequencies, and is nearly perpendicular to the local radial direction.
Recent work by Kontar et al. (2019) suggests that
shifts of this magnitude can be produced by propagation effects through the turbulent and anisotropic corona, and tend to be larger for sources closer to the limb.
Additionally the non-radial nature of the offset provides evidence for presence of significant non-radial anisotropy in the corona.

\subsection{Comparison of disk integrated solar flux density}
We compare the disk-integrated flux density, $S_{\nu}$, which is not sensitive to the distribution of flux density, obtained from MWA and FORWARD maps. 
A good match implies remarkable success on multiple fronts. 
On the observational front, it attests to the ability of MWA images to capture the bulk of solar emission and provides an independent verification of the techniques used for flux calibration \citep{Oberoi2017} and making $T_B$ maps \citep{Mohan2017}.
On the modeling front, it provides an independent verification of the veracity of both the physical state of the corona as determined by the MAS model, and the ability of FORWARD maps to predict a key radio observable based on the physical processes of thermal bremsstrahlung.

\begin{figure}
\begin{center}
\resizebox{75mm}{!}{
\includegraphics[trim={0.0cm 0cm 0.0cm 0.0cm},clip,scale=0.6]{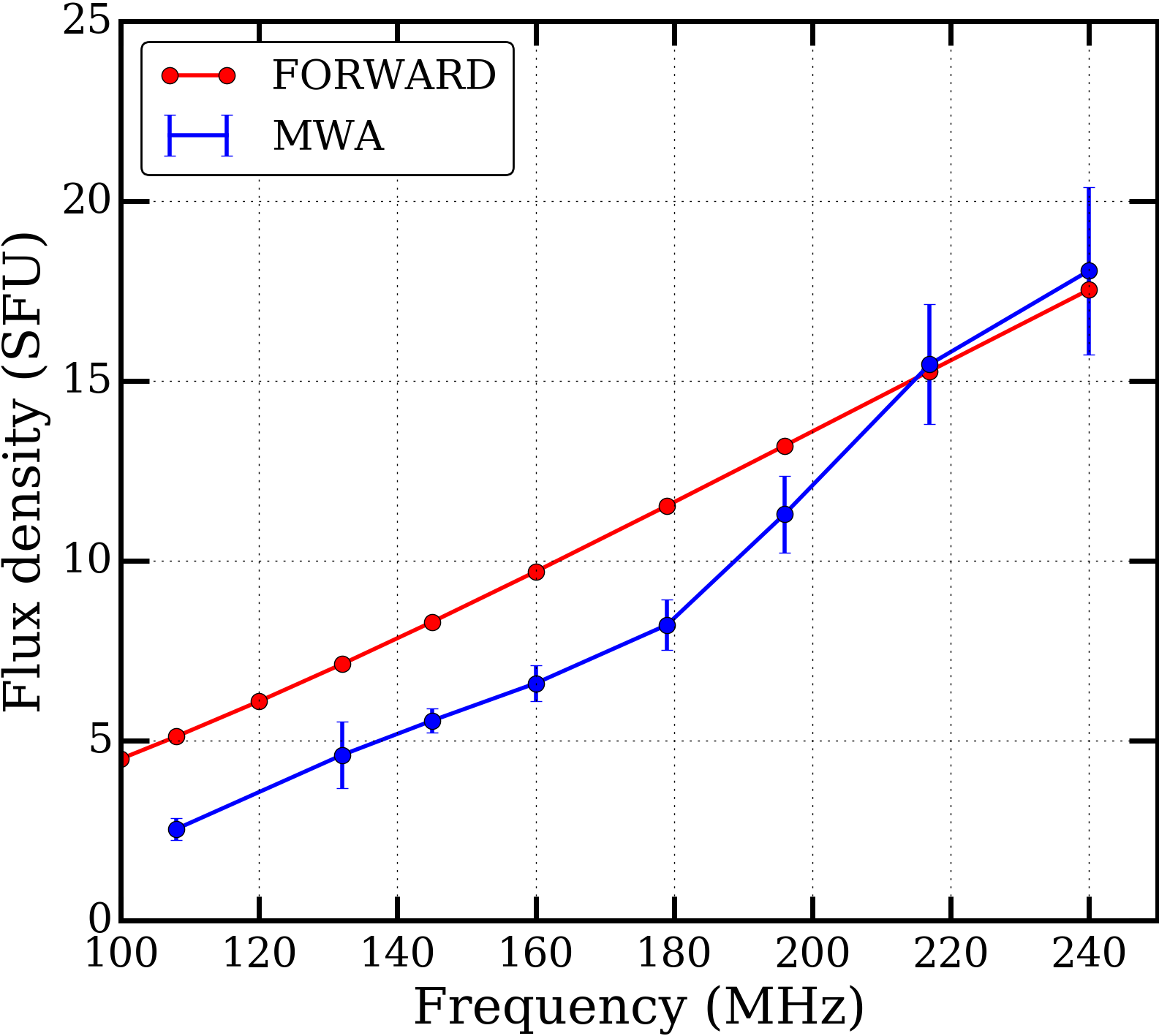}}
\end{center}
\caption{Solar disk integrated flux density for MWA and FORWARD maps as a function of frequency.}
\label{Fig:fwd_mwa_comp1}
\end{figure}

Figure \ref{Fig:fwd_mwa_comp1} shows a comparison of $S_{\nu}$ from FORWARD ($S_{\nu,\ FWD}$) and MWA ($S_{\nu,\ MWA}$)  maps. 
 For a given frequency, $S_{\nu, FWD}$ was computed using the following equation,
\begin{equation}
    S_{\nu, FWD} = \frac{2 k}{\lambda^2}\ \int_{Sun} T_{\nu, FWD}\ d\Omega,
\label{Eq:Snu-from-Tnu}
\end{equation}
where $k$ is the Boltzmann constant, $\lambda$ the wavelength of observation and the integration is carried out over a closed contour bounding the Sun.

$S_{\nu,\ MWA}$ is the sum of all of the flux density enclosed within the 5$\sigma$ contour in the radio map.
The errorbars on $S_{\nu,\ MWA}$ represent 1$\sigma$ of the flux density variation across the six baselines  used for flux calibration and over four min of time. 
$S_{\nu,\ FOR}$ and $S_{\nu,\ MWA}$ are remarkably consistent at the highest two frequencies. 
At lower frequencies $S_{\nu,\ MWA}$ is always significantly lower than $S_{\nu,\ FOR}$ and seems to settle at a value 2-4 SFU lower than $S_{\nu,\ FOR}$ at frequencies below 180 MHz.
This is very curious and implies the presence of a systematic effect which is leading to either an overestimate of $S_{\nu,\ FOR}$ or an under-estimate of $S_{\nu,\ MWA}$.
The excellent match between $S_{\nu,\ FOR}$ and $S_{\nu,\ MWA}$ at the two highest frequencies suggests that they could be used to explore the impacts of propagation effects.

\subsection{The FORWARD data cube}
\label{subsec:FWD-data-cube}
As preparation for understanding the differences between  $S_{\nu,\ FWD}$ and $S_{\nu,\ MWA}$ below 200 MHz, we first examine the 3D data FORWARD cube used for estimating $S_{\nu,\ FOR}$.
Just as in the photosphere, the coronal plasma density and temperature also show a significant variation. 
\begin{figure}
\begin{center}

\resizebox{70mm}{!}{\includegraphics[trim={0.0cm 0cm 0.0cm 0.0cm},clip,scale=0.13]{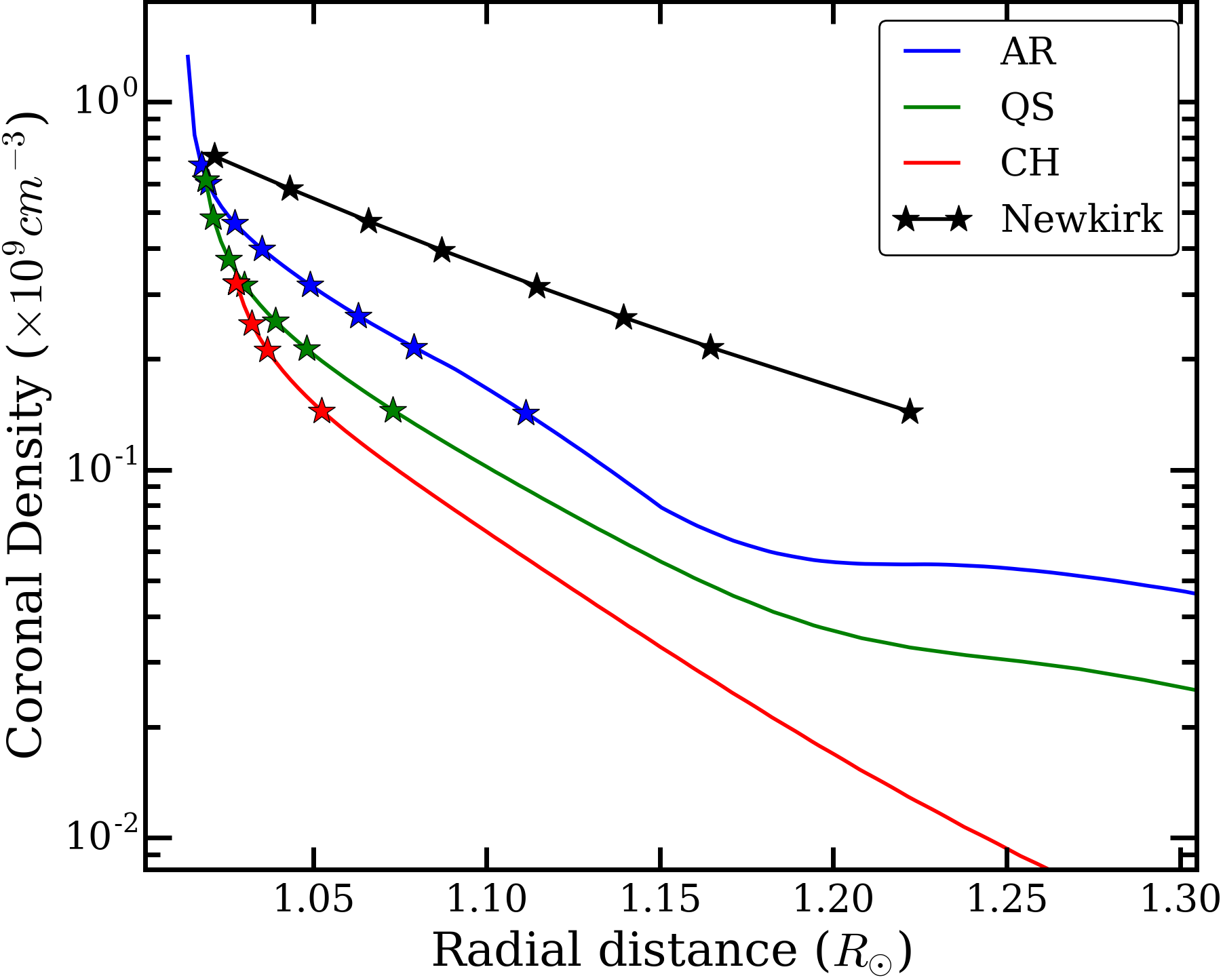}}

\resizebox{70mm}{!}{
\includegraphics[trim={0.0cm 0cm 0.0cm 0.0cm},clip,scale=0.13]{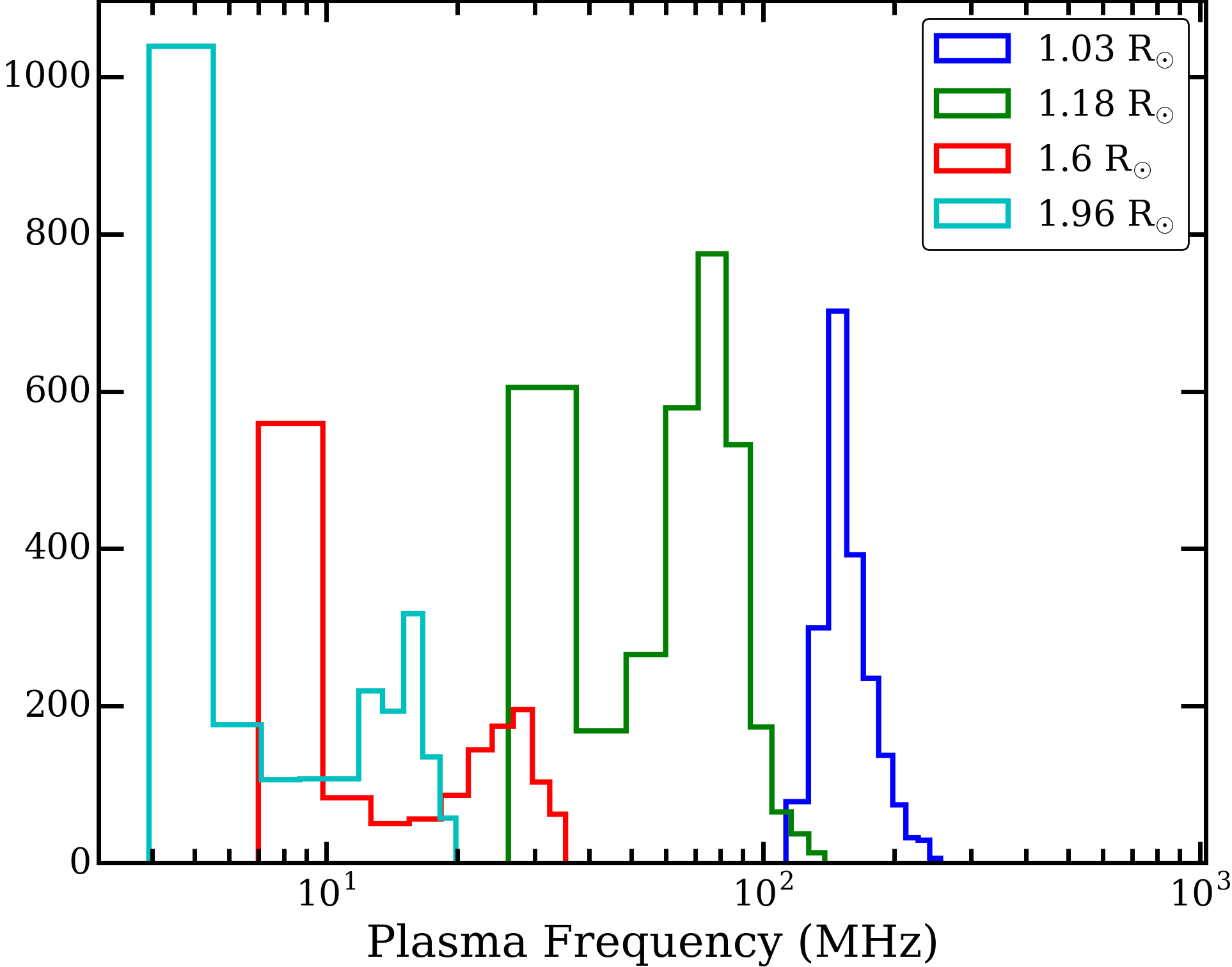}}
\end{center}
\caption{Top: Radial profile of coronal electron density for the active, quiet sun and coronal hole regions, and for the Newkirk model \citep{Newkirk1967}.
Bottom: The distribution of plasma frequencies at a few selected coronal heights. Note: the x-axis in lower panel is in log scale.}
\label{Fig:3D-FWD-model}
\end{figure}
The top panel of Fig. \ref{Fig:3D-FWD-model} shows the radial electron density profiles extracted from the 3D model used by FORWARD over regions illustrative of the active region (AR), a quiet sun (QS) region and the coronal hole (CH). 
The density profile predicted by the Newkirk model \citep{Newkirk1967} is also shown for comparison.
At the base of the corona, the electron densities are very similar for all of these regions and the Newkirk model, but diverge quickly with increasing radial distance.
We note that for all of the regions considered, the Newkirk model suggests significantly larger electron densities all through the corona. 
By 1.25 $R_{\odot}$, the FORWARD coronal hole electron density is about an order of magnitude smaller than that suggested by the Newkirk model.
The distribution of plasma frequencies in thin radial shells at a few selected coronal heights spanning the 1--2 $R_{\odot}$ range are shown in the bottom panel of Fig. \ref{Fig:3D-FWD-model}. 
 A variation of a factor of few in plasma frequency is seen at all heights.
The striking departures between the 3D FORWARD and the expectations from the Newkirk model drives home the limitations of spherically, or at least azimuthally, symmetric coronal electron density models.
The use of these models is quite common, e.g. to estimate coronal height of radio sources using the frequency of plasma emission.
It is, however, useful to keep in mind that this is driven solely by lack of better options; its results are only indicative in nature and should not be interpreted in great quantitative detail.

\subsection{Understanding the discrepancy between $T_{B,FWD}$ and $T_{B,MWA}$}
\label{subsec:missing_flux}

\begin{figure}
\begin{center}
\resizebox{75mm}{!}{\includegraphics[trim={0.0cm 0cm 0.0cm 0.0cm},clip,scale=0.13]{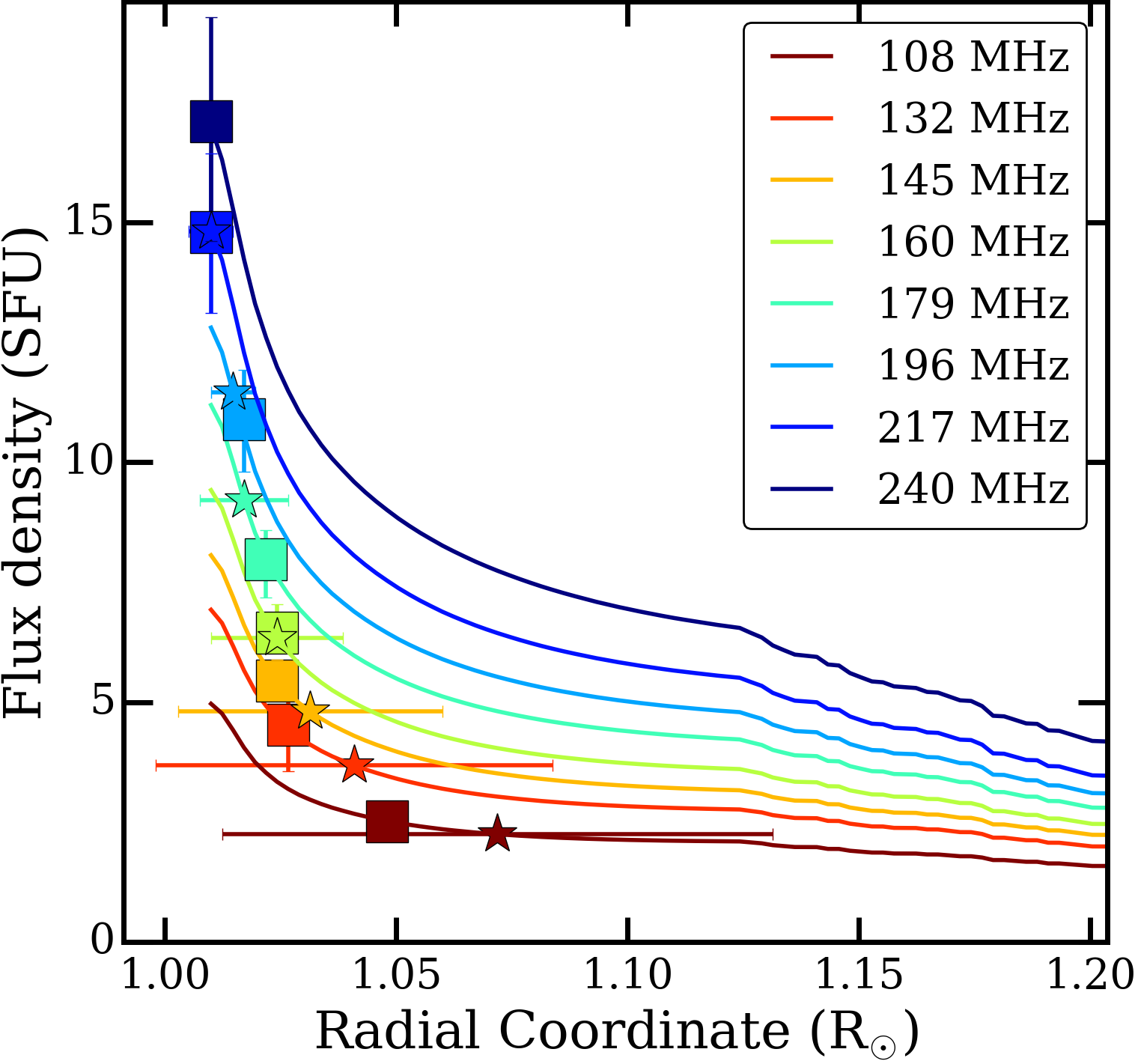}}
\end{center}
\caption{ The FORWARD flux density originating at different radial shells for MWA frequencies. The squares mark the disk-integrated MWA flux densities. The stars marks the  mean radial coordinate of the plasma frequency layer. The y-errorbars on squares corresponds to the errorbars on the integrated MWA flux density. The x-errorbars on stars are the RMS on the radial coordinate/shells in FORWARD density data cube corresponding to the plasma frequencies for each MWA frequency band. 
}
\label{Fig:LOS-flux}
\end{figure}

To make the task of comparing $S_{\nu, MWA}$ and $S_{\nu, FWD}$ more tracktable, we first compute an average coronal model and then do radiative transfer through it.
The average LoS is constructed by first dividing the 3D coronal model into thin radial shells (about 200 km in thickness) and computing the average physical properties for each shell. 
In section \ref{subsec:fwd_model}, we have simulated $T_B$ via radiative transfer from the base of the corona along a LoS (Eq. \ref{Eq:Tb}) all the way to the observer, by regarding the LoS to be comprising of numerous small segments.
Knowing the angular size of the source allows us to compute $S_{\nu}$ given $T_{\nu}$ (Eq. \ref{Eq:Snu-from-Tnu}).
Let $S_{i}$ denote the flux density measured from the $i^{th}$ line segment all the way to the observer.
Then $dS_{i} = S_{i}-S_{i+1}$ is the flux density contributed by the $i^{th}$ segment of the LoS.
$\sum_{j}^{N} dS_{i}$, where $N$ is the location of the observer, then provides an estimate of the flux density contributed by the part of the LoS beyond the $j^{th}$ radial segment. 
This quantity is shown in Fig. \ref{Fig:LOS-flux} for each of the MWA frequencies.
At any given radial distance, the plotted flux density represents the flux density contributed by the part of the LoS beyond that radial distance, all the way to the observer.
The filled box indicated the radial height at which the integrated $S_{\nu}$ along the average LoS equals the observed $S_{\nu, MWA}$.
This can be used as indicator of the effective coronal depth from which the LoS at a given frequency receives contribution. 
The $\star$ indicates the radial height at which the average plasma frequency of the coronal model equals the radio frequency of observation. 
As shown in Fig. \ref{Fig:3D-FWD-model}, there is large spread in the value of the plasma frequency in a given radial shell.
This spread is shown the error-bar associated with the $\star$.
This suggests that the thermal radiation at low frequencies comes not only from higher coronal heights, but also a much wider height range depending on coronal structures. 

That at low radio frequencies the observed $T_B$ for the sun is much lower than the MK coronal $T_e$ has been known for a while.
The commonly accepted explanation argues that scattering arising due to the presence of inhomogenities in the turbulent coronal medium effectively push the height of the critical layer, below which the LoS cannot penetrate, to much higher than that corresponding to the plasma frequency \citep[e.g.][]{Thejappa2008}.
As the FORWARD model does not incorporate any propagation effects, including scattering, the observed value of $S_{\nu, MWA}$ being less $S_{\nu, FWD}$ is consistent with this explanation.
However the rather sharp change from the two flux densities being consistent above 200 MHz and $S_{\nu, MWA}$ falling below $S_{\nu, FWD}$ suddenly at lower frequencies does seem somewhat surprising.

The MAS model is based on HMI extrapolated magnetic fields, and looses accuracy as one proceeds to higher coronal heights.
On the other hand, lower radio frequencies pick up larger fraction of their flux density contributions from exactly these higher coronal heights.
This suggests the possibility that perhaps the coronal model in use is not well suited for modeling higher coronal heights and is giving rise to systematic discrepancies between $S_{\nu, MWA}$ and $S_{\nu, FWD}$.

\section{Quantifying impacts  of propagation}
\label{Sec:quantifying-scattering}

Here we present our attempts to quantify impacts of propagation in terms of the fraction of radiation scattered and the estimated level of inhomogenities in the coronal medium.

\subsection{Scattered emission fraction}
We start with the assumption that that the FORWARD maps represent the true solar radio maps in absence of any propagation effects.
For 218 and 240 MHz, $S_{\nu, FWD}$ and $S_{\nu, MWA}$ are consistent and this makes a quantitative comparison interesting.

\begin{figure}
    \centering
    \resizebox{80mm}{!}{
\includegraphics[trim={0.0cm 0cm 0.0cm 0.0cm},clip,scale=0.33]{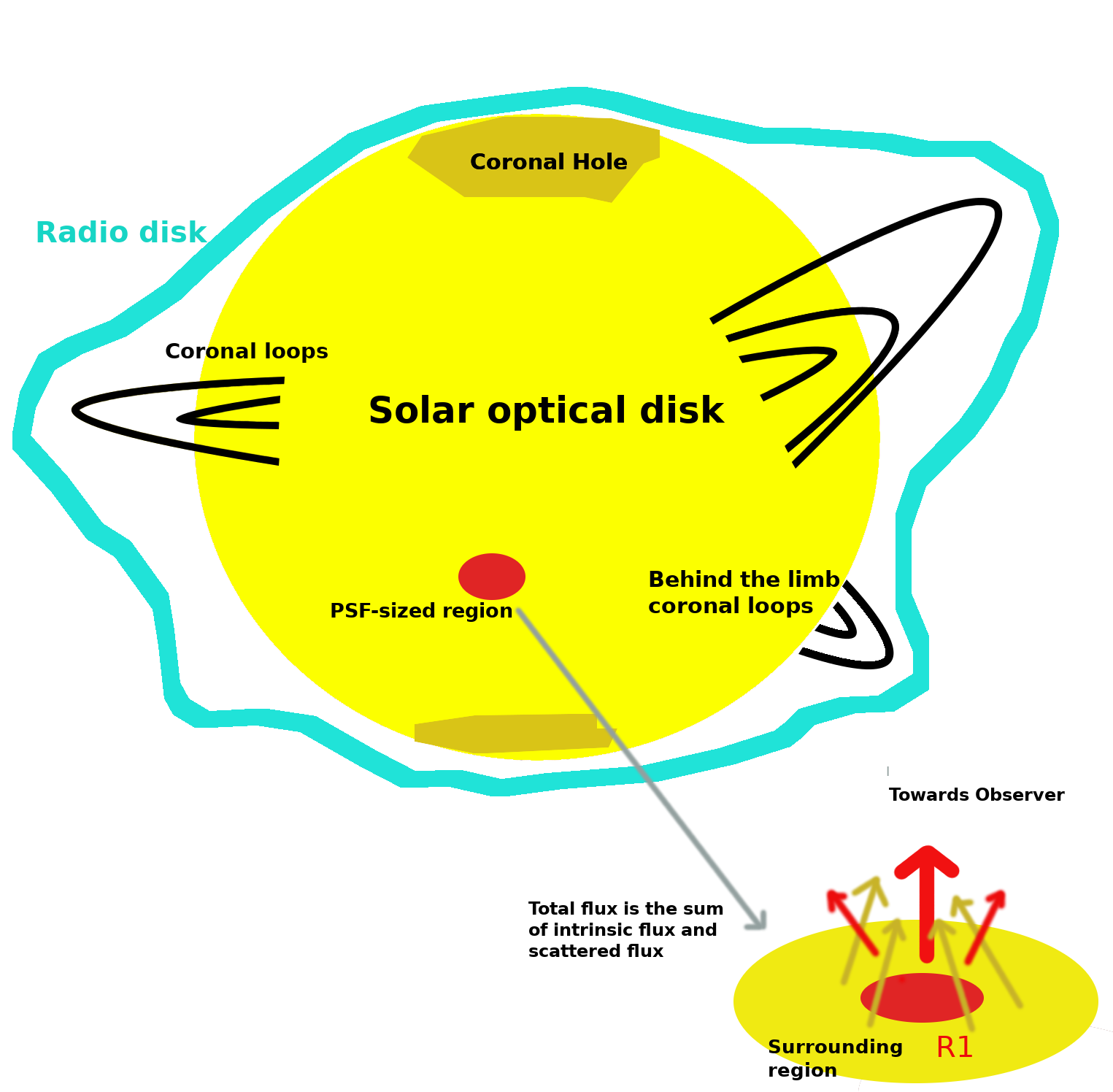}}
    \caption{Schematic showing the features of radio Sun and the presence of scattered flux in a PSF-sized region from its surrounding regions. Red region represents the region R1. The total flux within a PSF-sized region is a sum of its intrinsic flux (red arrow) and scattered flux (yellow arrows).}
    \label{Fig:sketch}
\end{figure}

The physical picture we model is shown in Fig.  \ref{Fig:sketch}. 
For a region R1, some fraction of the radiation emanating from it will be scattered out of the LoS to the observer, and similarly some radiation from neighboring regions will get scattered into the nominal LoS originating from R1.
$T_{B,R1,0}$ represents the $T_B$ of radiation emanating from region R1 in FORWARD maps (represented by $0$ in the subscript).
To keep the problem tractable, we represent the temperature of the neighboring region from which the radiation is scattered into the LoS by $T_{B,N,0}$, and it is arrived at by computing the mean $T_B$ over a somewhat arbitrarily defined region in the vicinity of R1. This region was chosen to be the one bounded by the 60$\%$ contour in FORWARD maps. This includes all the bright regions on the solar disk. It is reasonable to expect that the flux density scattered in to the line of sight will be dominated by the contribution from the brighter regions in the map.
We parameterize the strength of scattering by a parameter $\alpha$.
We assume that the nature of inohomogenities in the coronal regions covering regions is the same everywhere, and hence parameterized by the same $\alpha$.
The $T_B$ observed in the MWA maps $T_{B,R1,MWA}$ depends on $T_{B,R1,0}$, $T_{B,N,0}$ and $\alpha$, and can be modelled as
\begin{equation}
        T_{B,R1,MWA} = (1-\alpha)\ T_{B,R1,0} + \alpha\ T_{B,N,0}.
    \label{Eq:Scattering}
\end{equation}
Equation \ref{Eq:Scattering} can be rearranged to yield $\alpha$
\begin{equation}
      \alpha = \frac{T_{B,R1,MWA}-T_{B,R1,0}} {T_{B,N,0} - T_{B,R1,0}}\ .
      \label{Eq:alpha}
\end{equation}

$\alpha=0$ implies no scattering, strength of scattering increases with increasing values of $\alpha$, and $\alpha > 1$ is aphysical.

\begin{figure}
\begin{center}
\begin{tabular}{c}
\resizebox{80mm}{!}{\includegraphics[trim={0.0cm 0cm 0.0cm 0.0cm},clip,scale=0.32]{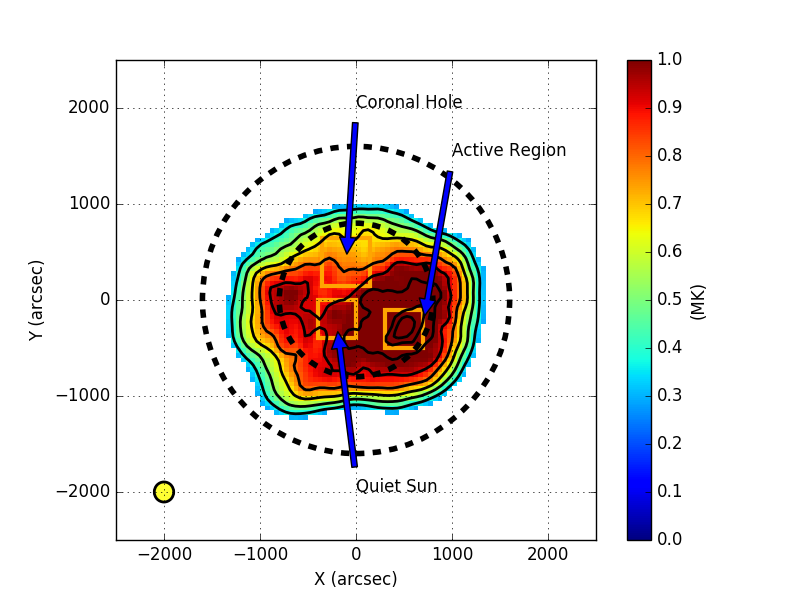}}
 \\
\resizebox{60mm}{!}{\includegraphics[trim={0.0cm 0cm 0.0cm 0.0cm},clip,scale=0.32]{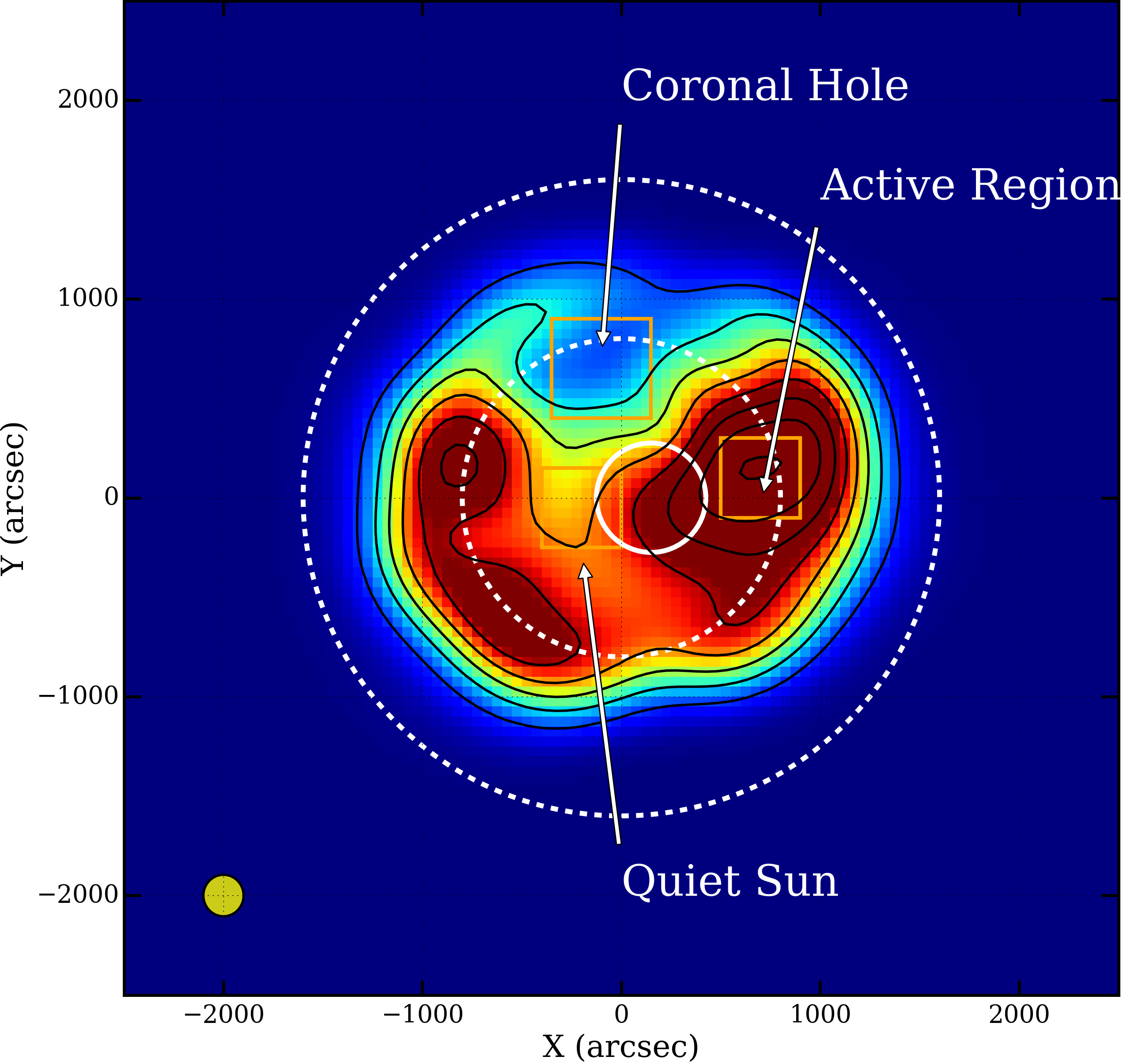}}\\
\end{tabular}
\end{center}
\caption{ Top Panel: 240 MHz MWA map along with the three representative regions for AR, QS and CH locations. Bottom Panel: FORWARD 240 MHz map, with the same regions marked. Note that the colorscales in both images are identical. The eight contours plotted are at 30, 40, 50, 60, 70, 80, 90 and 95 \% w.r.t the maximum in each of the maps. The white circle marks the area used for computation of inhomogeneity parameter in Section \ref{Sec:modelling}.}
\label{Fig:regions}
\end{figure}

In the MWA and FORWARD maps three distinct kinds of regions are easy to identify - bright active regions (AR), quiet sun regions (QS) and coronal hole regions (CH).
For the quantitative analysis three representative regions on the sun, shown in Fig. \ref{Fig:regions}, are chosen for each of these categories.
The regions corresponding to AR and CH are chosen from the FORWARD map, as they are easier to identify there, and the QS region is chosen to be near the centre of the disc. The size of the selected regions is in the range 3--4 $\times$ PSF size. The mean brightness temperature within the regions and the estimated values of $\alpha$ are listed in Table \ref{Tab:Tb_regions}. The variation in $T_B$ within these regions is found to be in the range 10--15\% (Table \ref{Tab:Tb_regions}). For $T_{B,FWD}$ this variation is used to compute an uncertainty on it.
Uncertainties in the values of $\alpha$ are computed by propagating the uncertainties in estimates of $T_{B,MWA}$ and $T_{B,FWD}$.

$\alpha_{AR}$ is $\sim 0.5$ for 217 and 240 MHz. 
However $\alpha_{QS}$ and $\alpha_{CH}$ are larger than unity,  though with large uncertainties.
This implies that for the QS and CH regions, either the model proposed in Eq. \ref{Eq:alpha} is insufficient for quantitatively modeling the strength of scattering, or some of the assumptions made are being violated, or both.

Assuming FORWARD models to be true representation of solar emission, we attempt to obtain a lower limit on the fraction of scattered emission, $f_{sc, R1}$ from a region $R1$ using 
\begin{equation}
    f_{sc, R1} =\frac{|T_{B,R1,MWA} - T_{B,R1,0}|}{T_{B,R1,0}}.
\end{equation}
A small value of $f_{sc}$ does not necessarily imply weak or low strength of scattering, but a larger value of $f_{sc}$ does imply an increasing strength of scattering.
We expect the observed $T_B$ for AR to decrease, while that for QS and CH to increase.
The values of $f_{sc}$ obtained are presented in Table \ref{Tab:Tb_regions}. 
The lowest estimated value of $f_{sc}$ is $\sim$0.14 for AR at 217 MHz, and the largest $\sim$3 at the same frequency for CH.
Such large values imply that scattering plays an important role in redistributing the flux density.

\begin{table}[]
    \centering
    \begin{tabular}{|c|c|c|c|c|}
\hline
Parameter & \multicolumn{2}{c|}{217 MHz}  &  \multicolumn{2}{c|}{240 MHz}
\\
\hline
\hline
T$_{B,AR,MWA}$&  \multicolumn{2}{c|}{1.40$\pm$0.15} &  \multicolumn{2}{c|}{1.23$\pm$0.16}   \\
T$_{B,QS,MWA}$& \multicolumn{2}{c|}{1.22$\pm$0.13} & \multicolumn{2}{c|}{1.07$\pm$0.15}  \\
T$_{B,CH,MWA}$& \multicolumn{2}{c|}{0.98$\pm$0.10} &  \multicolumn{2}{c|}{0.85$\pm$0.12} \\
\hline
T$_{B,AR,0}$& \multicolumn{2}{c|}{1.63$\pm$0.18}  & \multicolumn{2}{c|}{1.55$\pm$0.21}  \\
T$_{B,QS,0}$& \multicolumn{2}{c|}{0.79$\pm$0.08} & \multicolumn{2}{c|}{0.74$\pm$0.1} \\
T$_{B,CH,0}$& \multicolumn{2}{c|}{0.25$\pm$0.03} &  \multicolumn{2}{c|}{0.25$\pm$0.03}\\
\hline
T$_{B,N,0}$& \multicolumn{2}{c|}{1.11$\pm$0.11} &  \multicolumn{2}{c|}{0.97$\pm$0.14}\\
\hline
$\alpha_{AR}$& \multicolumn{2}{c|}{0.44$\pm$0.15} & \multicolumn{2}{c|}{0.54$\pm$0.21} \\
$\alpha_{QS}$& \multicolumn{2}{c|}{1.35$\pm$0.45}& \multicolumn{2}{c|}{1.42$\pm$0.61}\\
$\alpha_{CH}$& \multicolumn{2}{c|}{2.32$\pm$0.78}& \multicolumn{2}{c|}{2.59$\pm$1.06}\\
\hline
$f_{sc,AR}$ &  \multicolumn{2}{c|}{0.14 $\pm$ 0.05}  &  \multicolumn{2}{c|}{0.20 $\pm$ 0.08} \\
$f_{sc,QS}$  &  \multicolumn{2}{c|}{0.53 $\pm$ 0.18}  &  \multicolumn{2}{c|}{0.44 $\pm$ 0.18} \\
$f_{sc,CH}$  &  \multicolumn{2}{c|}{2.92 $\pm$ 0.99}  &  \multicolumn{2}{c|}{2.41 $\pm$ 0.99} \\
\hline
    \end{tabular}
    \caption{Brightness temperature, $\alpha$ and scattering fractional limit ($f_{sc}$) of various regions for MWA and FORWARD maps. Note that all brightness temperatures are in MK. The errorbars on $T_B$ are computed by propagating the errorbars of MWA fluxes, while region's RMS for FORWARD $T_B$s.}
    \label{Tab:Tb_regions}
\end{table}

\subsection{Level of coronal inhomogeneities}
\label{Sec:modelling}

\begin{figure}
\begin{center}
\resizebox{75mm}{!}{
\includegraphics[trim={0.0cm 0cm 0.0cm 0.0cm},clip,scale=0.13]{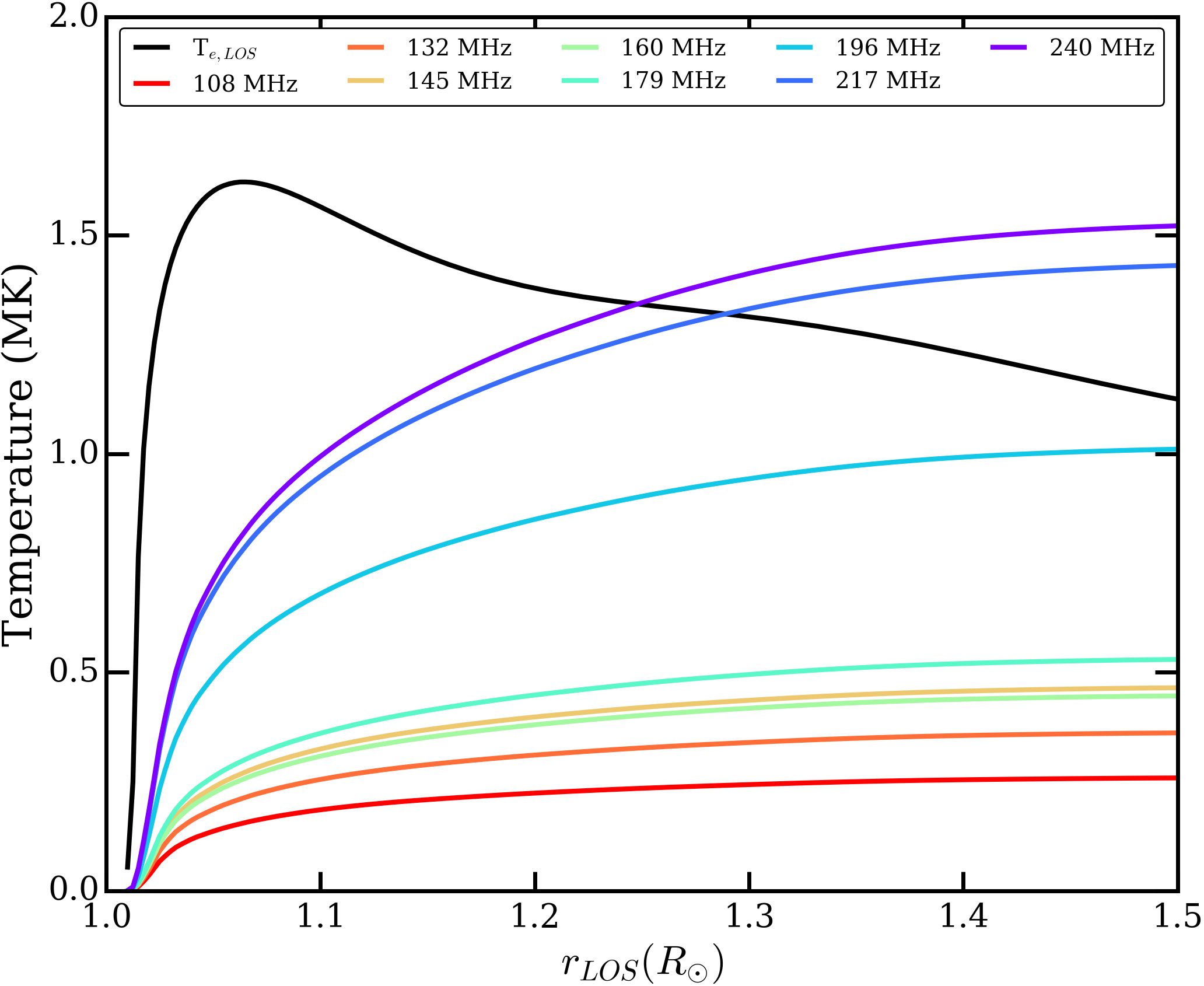}}
\end{center}
\caption{ LoS $T_{B}$ obtained by median averaging over a region of 11' around the centre in PoS. The black curve shows the plasma temperature along LoS for disk center.
}
\label{Fig:Tb_mean}
\end{figure}

The strength of scattering is well known to depend on the level of density inhomogenities, parameterized by $\epsilon = \frac {\Delta N}{N}$ , where $N$ is local average electron density and $\Delta N$, a measure of the departure from it.
Following earlier works, we assume that the inhomogenities have a Gaussian distribution with a characteristic correlation length scale $h$ \citep{Chandrasekhar1952,Hollweg1968,Steinberg1971,2018ApJ...868...79C,Kontar2019}.
For such a medium and small-angle scattering, the scattering optical depth, $\tau_{SC}$, along a LoS at a frequency $f$ is given by
\begin{equation}
    \tau_{sc} (r) = \int_{r}^{\infty} \frac{\sqrt{\pi}f_{pe}^{4}(r)}{2(f^2 - f_{pe}^{2} (r))^2} \frac{\epsilon^2}{h} dr,
    \label{Eq:tau_sc}
\end{equation}
where $f_{pe}$ is the electron plasma frequency and $r$ the radial coordinate.
To take the effect of scattering into account, we modify the radiative transfer equation (Eq. \ref{Eq:Tb}) as follows:

\begin{equation}
\begin{array}{r@{}l}
    T_{B,i} = & T_{B,i-1}  e^{-(d\tau_i + d\tau_{sc,i})}\ +\ T_{e,i}(1-e^{-d\tau_i}) \nonumber\ \\
              & +\ \frac{1}{\gamma}\ T_{B,mean,i} (1-e^{-d\tau_{sc,i}}). 
\end{array}\label{Eq:Tb_sc}
\end{equation}

Here $T_{B,i}$ represents the radiation temperature at the $i^{th}$ segment of the LoS, $T_{B,mean,i}$ that corresponding to the mean brightness temperature in the neighbouring regions, and $d\tau_{sc, i}$ the scattering optical depth for the $i^{th}$ LoS segment.
$\gamma$ is the ratio of $S_{\nu, FWD}$ to $S_{\nu, MWA}$ and is listed in Table \ref{Tab:temp}. 
The first term now also includes a parameterization of the fraction of incident brightness temperature removed from the LoS due to scattering. 
The second term is as before, and the third term represents the contribution scattered in to the LoS from the neighboring regions.
Though it is not expected to be the case, for lack of a better option, $\epsilon$  and $h$ are assumed to remain constant along the LoS.

 Refraction can have a considerable impact on the rays propagation out through the corona. 
Even for the simplest case of an isothermal spherically symmetric corona, the apparent shift of sources towards the disk centre due to refraction can lead to a change in the observed $T_B$.
This effect is the smallest for LoS close to the centre of the solar disk and is neglected here for simplicity. 
We assume that the $T_{B, FWD}$ has been modified only due to scattering to result in $T_{B, MWA}$.
We also need to choose a region over which $T_{e,mean,i}$, needed in Eq. \ref{Eq:Tb_sc}, can be computed.
Somewhat arbitrarily, we use a region of diameter of $11'$ around the central LoS (shifted towards brighter region) to compute $T_{e,mean,i}$ (Fig. \ref{Fig:regions}). 
This is justified because the flux scattered in will be dominated by the brighter regions in the neighborhood.
Figure \ref{Fig:Tb_mean} shows the variation in $T_{B,mean}$ along the LoS for the frequencies of interest.

\begin{figure*}
\begin{center}
\begin{tabular}{ccc}
\resizebox{49mm}{!}{
\includegraphics[trim={0.0cm 0cm 0.0cm 0.0cm},clip,scale=0.32]{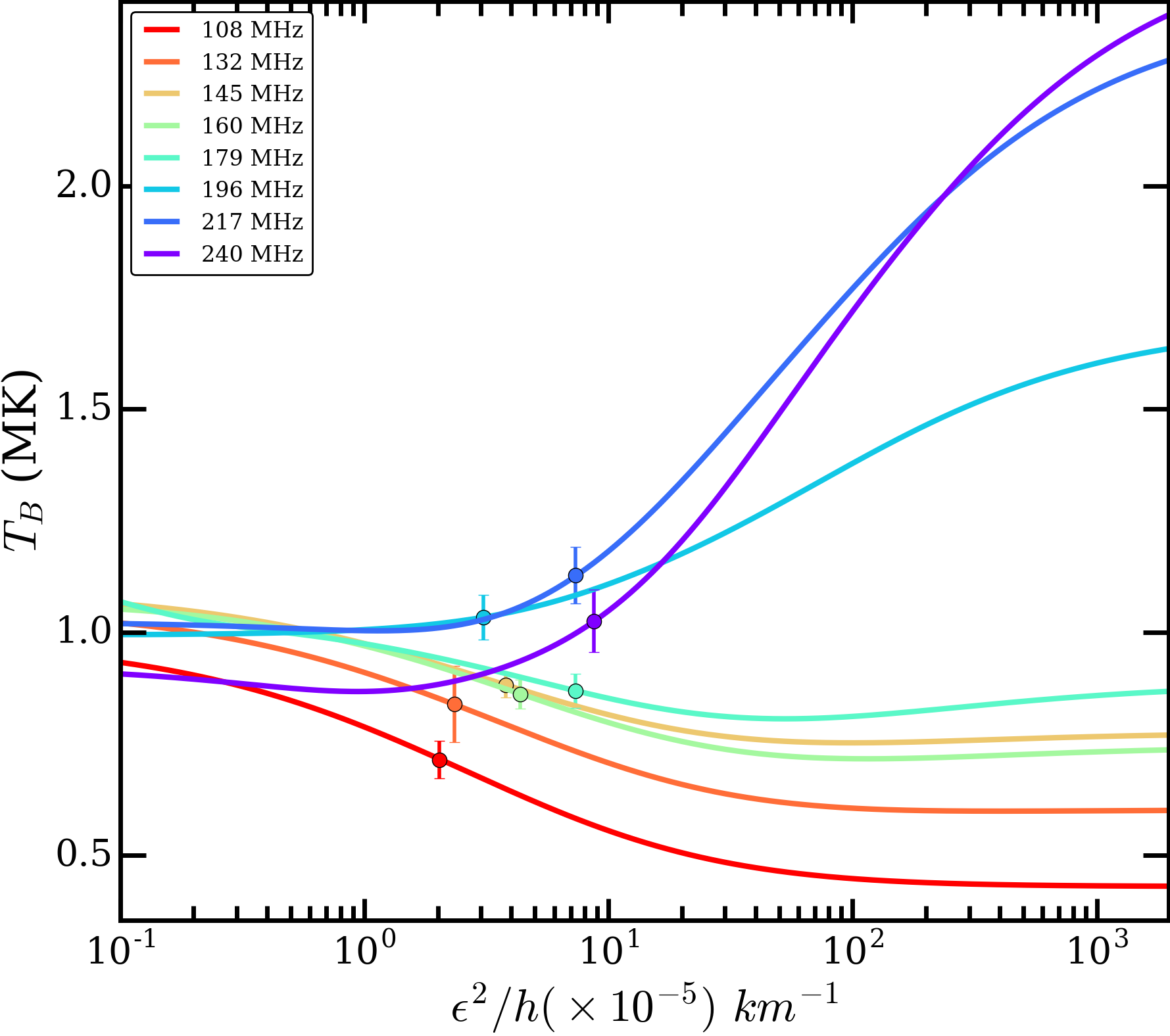}}
 &
\resizebox{50mm}{!}{
\includegraphics[trim={0.0cm 0cm 0.0cm 0.0cm},clip,scale=0.32]{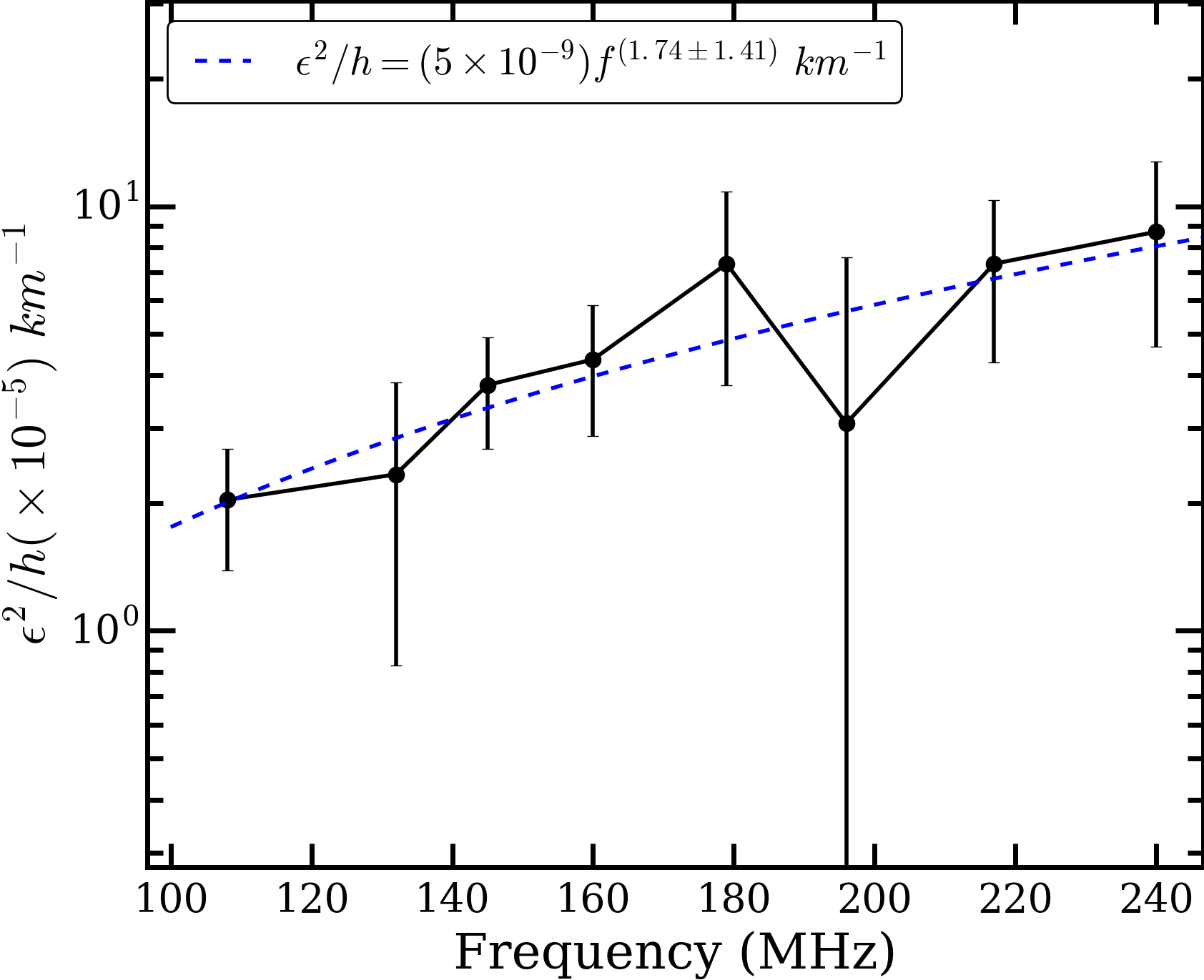}}
 &
\resizebox{50mm}{!}{
\includegraphics[trim={0.0cm 0cm 0.0cm 0.0cm},clip,scale=0.32]{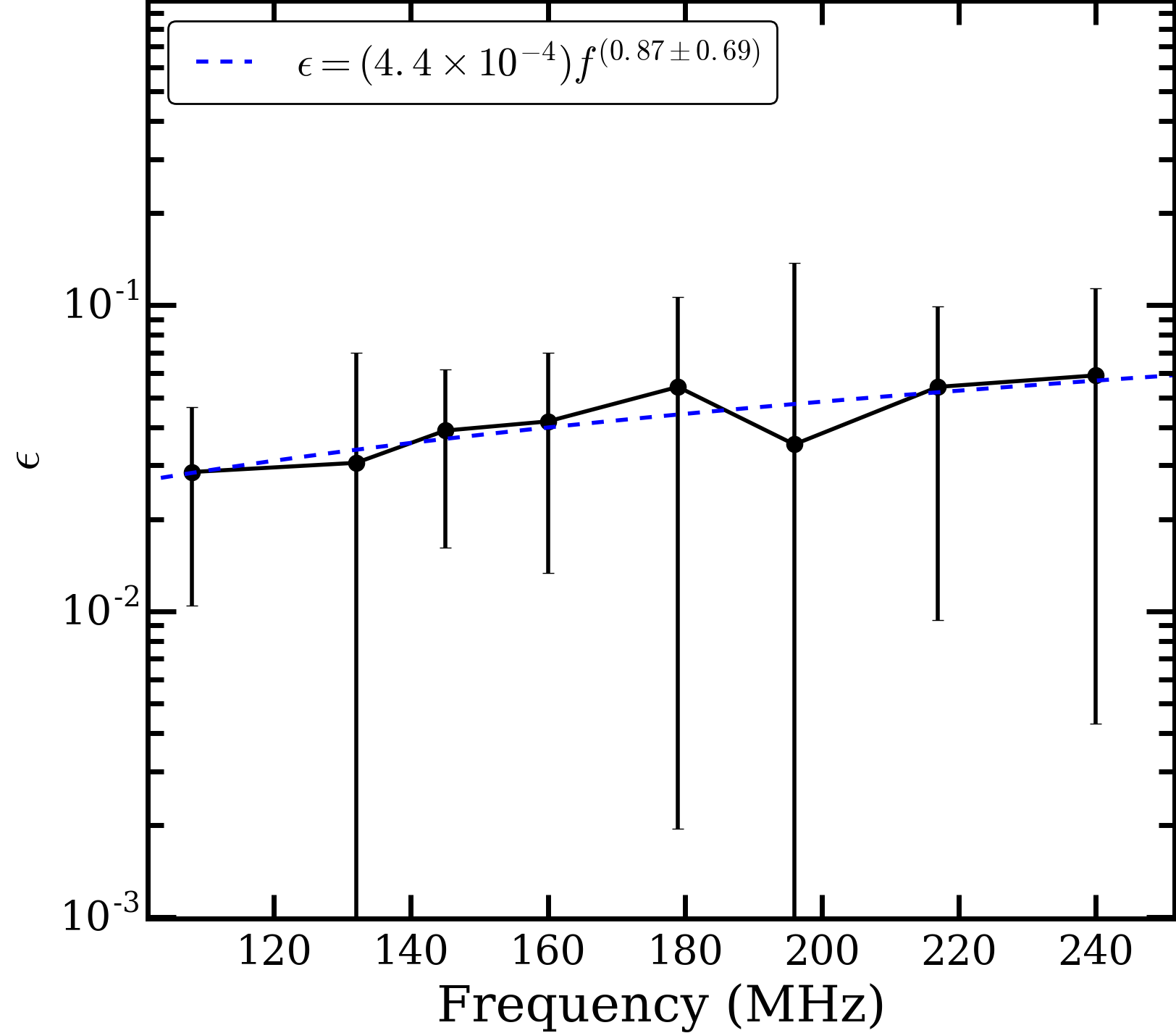}}
\end{tabular}
\end{center}
\caption{ Left panel: The brightness temperature obtained by radiative transfer for disk-centre LoS by varying $\epsilon^2/h$. The points on the curve are the MWA $T_B$ for disk-centre LoS. Middle panel: The optimised values of $\epsilon^2/h$ for MWA frequency bands. The errorbars are the propagated uncertainties in MWA fluxes to brightness temperatures and $\epsilon^2/h$. The dashed line shows the power-law fit. Right panel: The optimised values of $\epsilon$ for MWA frequency bands. Note that here $h\approx 40$ km. The dashed line shows the power-law fit to the $\epsilon$ datapoints.}
\label{Fig:ep2byh}
\end{figure*}

Using the stated assumptions, one can relate $T_{B,MWA}$ to $T_{B,FWD}$ for the central LoS with the impact of scattering parameterized by a single parameter, $\epsilon^2/ h$.
For each of the MWA frequencies, the left panel of Fig. \ref{Fig:ep2byh} shows the expected value of the observed $T_{B}$ for the central LoS as a function of $\epsilon^2/h$. 
The observed values of $T_{B,MWA}$ for each of the observing frequencies are marked, along with their uncertainties, and for all of them $\epsilon^2/h$ lies in the range 1--10$\times 10^{-5}\ km^{-1}$. 
For the lower frequencies, with increasing $\epsilon^2/h$, $T_B$ tends to decrease and settle down to low values, while at higher frequencies, it tends to continue to increase to larger values.
The uncertainty in $T_B$ have been translated to that in $\epsilon^2/h$ in the middle panel of Fig. \ref{Fig:ep2byh}.
Averaging across frequency yields $\epsilon^2/h \approx (4.9\pm2.4)\times 10^{-5}\ km^{-1}$, the RMS across the frequencies is used as the uncertainty. 
The right panel of Fig. \ref{Fig:ep2byh} shows the values of $\epsilon$ arrived at by assuming $h=40$ km \citep{Steinberg1971}, with an average value of $\epsilon = (4.28\pm1.09)\%$. 

\subsection{Uncertainties}

In the current analysis, the major sources of uncertainties can be categorized into those arising from observations and those from limitations of the modeling formalism.

\subsubsection{Observational uncertainties}

The primary sources of observational uncertainty are the absolute flux calibration procedure prescribed by \citet{Oberoi2017}, and those arising from imaging calibration and deconvolution process.
The latter are quantified by the RMS measured in the residual radio maps at the end of the deconvolution process in regions far from the Sun.
It turns out that the uncertainties are dominated by those arising from the absolute flux calibration procedure which typically lie in the range  8\% to 21\%.
These uncertainties are reflected in the error bars in $T_B$ shows in Fig. \ref{Fig:ep2byh}.

Ionospheric propagation effects are also well known to give rise to distortions in the image plane at meter wavelengths \citep[e.g.][]{Loi2015}.
However for an instrument with a small footprint like the MWA, the bulk of these distortions are limited to large scale refractive shifts.
The differential shifts across an object of the size of solar disk ($< 1^{\circ}$) are expected to be much smaller than the PSF for these observation, and can justifiably be ignored here.

\subsubsection{Limitations of models}

Modeling of radio emission from the coronal plasma requires detailed information of the electron density and temperature distributions.
For these we rely on the PSIMAS model.
PSIMAS provides a robust estimate of large scale stationary structures.
However it does not have information about the variations at small spatial scales, which play the defining role in scattering.
Also the measurements on which PSIMAS relies progressively loose their SNR with increasing radial heights, leading to decreasing accuracies with height.
We suspect this plays a role in the FORWARD model overestimating the radio flux density at lower frequencies, which preferentially sample higher coronal heights.
In addition, the inability of the FORWARD model to take the effects of scattering and refraction into account has already been discussed in Sec. \ref{subsec:fwd_model}.

The FORWARD model used to predict the radio maps from the physical description provided by PSIMAS only models the free-free emission.
Large scale magnetic field distribution is available from PSIMAS, but is not made use of by FORWARD to predict the gyroresonance emissions. 
Given the very low expected strength of this emission, it is perfectly reasonable to ignore it.
Similarly, the gyrosynchrotron emission is also ignored.

\section{Discussion \& Summary}
\label{Sec:discssion}

We present a novel approach for studying propagation effects in quiet Sun corona based on a quantitative comparison between a data-driven empirical model for thermal bremsstrahlung emission and observed solar maps at meter wavelengths. Both of the ingredients needed for such a study have become available only comparatively recently. 
From having synthesis imaging maps using data spanning hours at a few discreet frequencies \citep[e.g.][]{Alissandrakis1985, Mercier2015, Vocks2018}, we have come to being able to make
high fidelity maps of the quiet sun over essentially snapshot observations with a much denser spectral sampling.
The ability to make realistic coronal models has also improved remarkably recently, and with the availability of tools like FORWARD, they have also become very accessible \citep[e.g.][]{Zucaa2014, Gibson2015, Gibson2016}.
This approach can prove to be a powerful tool for building the a quantitative sense for effects of scattering and perhaps help identify the deficiencies of the models used.
An agreement between the integrated modelled and observed solar flux densities demonstrate the reliability of the model at frequencies above 200 MHz, while the discrepancy at the lower frequencies suggests need for improvements in the models. 
Among the reasons for the latter could be the absence of propagation effects in the model itself.

The MAS model and the corresponding thermal bremsstrahlung solar maps show a diversity of solar features. 
Associations between solar surface features, like active regions and coronal holes, and features in radio maps have commonly been observed \cite[e.g.][]{Alissandrakis1996,Lantos1992,Lantos1999, Mercier2015, Rahman2019} and are also seen in these data. 
In addition, we also observe a hint of limb brightening (Fig. \ref{Fig:108MHz}).
The large variations are seen in $n_e$ in MAS models at any radial height (Fig. \ref{Fig:3D-FWD-model}).
This challenges the notion of a simple radial dependence of density \cite[e.g.][]{Newkirk1967, Saito1977}, and the idea that radiation at a given frequency can essentially be regarded at originating a specific coronal height where the local plasma frequency equals the observing frequency.

Scattering and refractive effects redistribute the radio emission on the solar disk, significantly changing the appearance of the radio sun \citep[e.g.][]{Aubier1971,Rahman2019,Kontar2019}.
Refraction arises from large scale and gradual density gradients and scattering from small scale density fluctuations. 
Coronal refraction tends to shift sources closer to the disk centre, while scattering, not only increases the apparent observed size, it also tends to shift the apparent location of sources farther from the disk centre. 
So, in this sense, they tend to work in opposite ways.
Anisotropic scattering from non-radial coronal structures can give rise to non-radial shifts in the apparent locations of sources.
The peak of the emission associated with the active regions shifts by 8'--11' between the observed and modelled maps (Table \ref{Tab:temp}).
The angular size of the sun in MWA images is found to be larger than that in FORWARD images by 25--30\% (Table \ref{Tab:size}).
Even though our observing frequencies span more than a factor of 2, the difference between the observed and the modeled size remains roughly constant.
The absence of the expected frequency dependence ($\propto 1/\nu^2$) in the increase in the size of the scattered disk implies that the observed size is not dominated by scattering affects alone.
A possible reason could be that the expected increase in the angular size is being effectively countered by the refractive effects.
Additionally, the expected frequency dependence holds only for the conventional view of magnetic fields being largely radial. 
Large departures of magnetic fields from being radial can also have a significant impact on the observed sizes.

As already mentioned, scattering leads to a shift in the observed location of the sources closer to the disk centre. 
We examine this shift by comparing the locations of the brightest source in the MWA maps to that in the FORWARD maps, assuming them to correspond to the same physical source.
This source is associated with the active region and is closer to the limb than the disk centre.
This observed shift is in the range 8'--11' and is clearly non-radial (Fig. \ref{Fig:size}), suggesting the presence of non-radial structures in the coronal medium. 
In general, the key element giving rise to the anisotropy is the presence of magnetic fields. 
Such anisotropy is expected in the vicinity of active regions which are populated by closed magnetic field loops.
The scattering characteristics are expected to vary considerably across the solar disk depending upon the nature of surface features and the associated magnetic fields. 

Assuming scattering to be the dominant effect in the observed maps, we present rough quantitative estimates of the fraction of scattered radiation using simple models.
We estimate that for coronal holes, the observed $T_B$ could be dominated by a factor between 2 and 3 by the radiation which has been scattered in to these lines of sight from the neighboring regions (Table \ref{Tab:Tb_regions}); while for active regions the up to half the flux density might be scattered out of these lines of sight.
Thus the observed $T_B$ depends not only on the intrinsic $T_B$ of the region from where the radiation is originating, but also on the $T_B$ of the neighboring regions. 
Recently successful detection of ubiquitous impulsive emissions of flux densities down to $\sim$mSFU levels have been reported \citep{Mondal2020b}.
In this context, we note that the significant smearing in the image caused by scattering will make it harder to detect intrinsically low level variations in solar radio images. Though on the other hand, studies of time profiles of such emissions and their observed shapes and sizes is likely provide interesting information about the nature of scattering and coronal inhomogenities.

The coronal density inhomogenities arise due to turbulence and it is interesting to quantify the level of density fluctuations, $\epsilon$.
\citet{Thejappa1994} estimated an $\epsilon \approx 12\%$, while lower values $\approx 1\%$ have been reported using the statistical ray-tracing simulations of \cite[e.g.][]{Steinberg1971, Krupar2018, Krupar2020} at decameter wavelengths. 
\citet{Mohan2019} estimated $\epsilon$ to be in the range 1--2\% at coronal heights around 1.4 $R_{\odot}$ using studies of weak type III radio bursts.
Using the completely independent technique of transcoronal spacecraft radio sounding observations \citet{Wexler2019} report similar values of $\epsilon$ at low coronal heights which increase to about 10\% by 14 $R_{\odot}$.
We estimate the inhomogeneity parameter  $<\epsilon^2/h>$ for the central LoS to be $\approx 4.9\pm2.4 \times 10^{-5}\ km^{-1}$. 
Using $h \approx$40 km \citep{Steinberg1971}, leads to $\epsilon \approx 4.28\pm1.09 \%$, consistent with earlier estimates.
Variations in the estimated values of $\epsilon$ could reflect the diversity of coronal physical conditions.
All measurements of $\epsilon$ have large uncertainties (Fig. \ref{Fig:ep2byh}, last panel).
The coronal inhomogenities observed close to the Sun advect away into the solar wind and evolve, and has routinely been estimated using remote sensing techniques like Interplanetary Scintillation. 
Recently the {\it in-situ} measurements by Parker Solar Probe have yielded $\epsilon \approx$ 6--7\% \citep{Krupar2020}, implying that $\epsilon$ has not evolved significantly even out to much larger solar distances.

We have investigated the role of propagation effects, especially scattering, using a detailed comparison of the observed and simulated meterwave radio maps.
This first attempt has already illustrated the magnitude and the complexities in the manner in which the solar radio emission gets modified by the propagation effects.
This has been made possible by the confluence of two independent developments - the availability of sufficiently high quality radio imaging with the new generation interferometers like the MWA; and that of robust data driven coronal models which can provide predictions of radio observables, like FORWARD.
The obvious omission is a tool to model the propagation effects. 
In the process of relating the FORWARD predictions to the observed MWA radio maps, such a tool can independently constrain properties like large scale density gradients as well as statistical descriptions of coronal turbulence and inhomogenities, and yield interesting physics.
This would also be the next logical step, building on the recent developments in developing a 3D stochastic description of coronal propagation.
We hope that this work will engender interest in the larger community to pursue this approach.


\acknowledgments

This scientific work makes use of the Murchison Radio-astronomy Observatory (MRO), operated by the Commonwealth Scientific and Industrial Research Organisation (CSIRO).
We acknowledge the Wajarri Yamatji people as the traditional owners of the Observatory site. 
Support for the operation of the MWA is provided by the Australian Government's National Collaborative Research Infrastructure Strategy (NCRIS), under a contract to Curtin University administered by Astronomy Australia Limited. We acknowledge the Pawsey Supercomputing Centre, which is supported by the Western Australian and Australian Governments. 
The SDO is a National Aeronautics and Space Administration (NASA) spacecraft, and we acknowledge the AIA  science team for providing open access to data and software. 
This research has also made use of NASA's Astrophysics Data System (ADS). 
RS acknowledges support of the Swiss National Foundation, under grant 200021\_175832, and of Marina Battaglia, FHNW, Windish, Switzerland.
DO acknowledges support of the Department of Atomic Energy, Government of India, under the project no. 12-R\&D-TFR-5.02-0700.

%





\bibliographystyle{aasjournal}
\bibliography{manuscript}{}



\end{document}